\documentclass[%
 reprint,
 amsmath,amssymb,
 aip,
]{revtex4-1}

\usepackage{listings}
\usepackage{comment}
\usepackage{xcolor}
\usepackage{graphicx}
\usepackage{subcaption}
\usepackage{dcolumn}
\usepackage{bm}
\usepackage{siunitx}

\newcommand{\secref}{Section~\ref}
\newcommand{\appref}{Appendix~\ref}
\newcommand{\figref}[2][{}]{\figurename~\ref{#2}#1}

\newcommand*\dif{\mathop{}\!\mathrm{d}}
\newcommand*\rme{\mathrm{e}}
\newcommand*\rmi{\mathrm{i}}

\begin{document}


\title{Machine learning characterisation of Alfv\'{e}nic and sub-Alfv\'{e}nic chirping and correlation with fast ion loss at NSTX}
\author{B. J. Q. Woods}
	\affiliation{School of Mathematics, University of Leeds, Woodhouse Lane, Leeds, LS2 9JT, United Kingdom}
	\affiliation{Department of Physics, York Plasma Institute, University of York, Heslington, York, YO10 5DD, United Kingdom}
\author{V. N. Duarte}
	\affiliation{Princeton Plasma Physics Laboratory, Princeton University, Princeton, NJ 08543, United States of America}
\author{E. D. Fredrickson}
	\affiliation{Princeton Plasma Physics Laboratory, Princeton University, Princeton, NJ 08543, United States of America}
\author{N. N. Gorelenkov}
	\affiliation{Princeton Plasma Physics Laboratory, Princeton University, Princeton, NJ 08543, United States of America}
\author{M. Podest\`{a}}
	\affiliation{Princeton Plasma Physics Laboratory, Princeton University, Princeton, NJ 08543, United States of America}
\author{R. G. L. Vann}
	\affiliation{Department of Physics, York Plasma Institute, University of York, Heslington, York, YO10 5DD, United Kingdom}

\date{\today}

\begin{abstract}
Abrupt large events in the Alfv\'{e}nic and sub-Alfv\'{e}nic frequency bands in tokamaks are typically correlated with increased fast ion loss. Here, machine learning is used to speed up the laborious process of characterizing the behaviour of magnetic perturbations from corresponding frequency spectrograms that are typically identified by humans. Analysis allows for comparison between different mode character (such as quiescent, fixed-frequency, chirping, avalanching) and plasma parameters obtained from the TRANSP code such as the ratio of the neutral beam injection (NBI) velocity and the Alfv\'{e}n velocity ($v_{\textrm{inj.}}/v_{\textrm{A}}$), the $q$-profile, and the ratio of the neutral beam beta and the total plasma beta ($\beta_{\textrm{beam},i}/\beta$). In agreement with previous work by Fredrickson \emph{et al.} [Nucl. Fusion 2014, 54 093007], we find correlation between $\beta_{\textrm{beam},i}$ and mode character. In addition, previously unknown correlations are found between moments of the spectrograms and mode character. Character transition from quiescent to non-quiescent behaviour for magnetic fluctuations in the 50 - 200 kHz frequency band is observed along the boundary $v_{\varphi} \lessapprox \frac{1}{4}(v_{\textrm{inj.}} - 3v_{\textrm{A}})$ where $v_{\varphi}$ is the rotation velocity.
\end{abstract}

\maketitle



\section{\label{sec:int}Introduction}
Fast ion instabilities correlated with increased Alfv\'{e}nic activity could prove to be a serious limitation to the nominal ITER performance;\cite{heidbrink2008basic} these wave-particle instabilities transfer free energy between the particle distribution function and plasma waves in the system, in certain cases leading to sudden degradation of plasma performance and energy confinement. Abrupt large events (ALEs or mode avalanches) are characterised by magnetic perturbations in the plasma undergoing very rapid frequency change (`chirping') across a broadband of frequencies, and are directly correlated with large energetic particle (EP) losses \cite{fredrickson2006fast,bierwage2018simulations}; understanding the parametric dependencies on these losses is essential for good plasma performance. These events are sudden and highly distinguishable from the frequency behaviour at times when ALEs do not occur, and are typically observed in the upper part of the kink/tearing/fishbone (KTF) frequency band ($\sim$ \SIrange{1}{30}{\kilo\hertz}) and the lower part of the toroidal Alfv\'{e}n eigenmode (TAE) frequency band ($\sim$  \SIrange{50}{200}{\kilo\hertz}). \cite{podesta2009experimental}

Tokamaks feature a high-dimensional parameter space (ion temperature, magnetic flux density, ion number density, etc.) with large operational domains. For certain plasma parameters, stability transition has been shown to suddenly occur over small regions of parameter space, as is observed with the L-H mode transition \cite{itoh1988model} and edge localized mode (ELM) crashes \cite{connor1998edge}; accordingly, certain regions of parameter space have boundaries between different states of plasma stability. However, the transition that leads to ALEs is not fully understood; previous work has shown correlation between fast ion beta and neutral beam injection (NBI) beam energy \cite{fredrickson2014parametric}, but a large area of parameter space still requires analysis.

\begin{figure}[h!]
	\centering
	\includegraphics[width=0.45\textwidth]{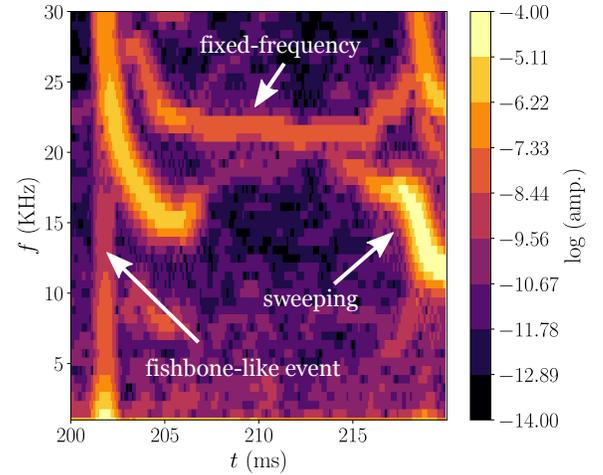}
	\caption{Frequency spectrogram for magnetic fluctuations on NSTX in the \SIrange{1}{30}{\kilo\hertz} (kink/tearing/fishbone) band, \SIrange{200}{220}{\milli\second} after the beginning of shot 139317. 3 types of mode character are observed: fixed-frequency eigenmodes, sweeping eigenmodes, and a fishbone-like event.}
	\label{fig:KTF}
\end{figure}

Previous work has highlighted the relationship between microturbulence, stochastic effects and fast ion loss \cite{duarte2017prediction,woods2018stochastic}, showing increased suppression of Alfv\'{e}nic chirping as a function of microturbulence; theoretically microturbulence is treated as a stochastic term in the pitch-angle scattering rate, while experimentally this can be heuristically inferred via the ion thermal conductivity $\chi_i$. Additionally, work by Van Zeeland \emph{et al.} has shown that Alfv\'{e}nic chirping is increased in negative triangularity plasmas, where the microturbulence is expected to be lower.\cite{vanzeeland2019alfven} However, there are other correlations which may yet be undiscovered.

Unfortunately, as one might expect in high-dimensional spaces, observing and predicting correlations becomes increasingly more difficult. With traditional computational analysis, there is no way to circumvent the amount of time required to perform human categorisation; a simple back-of-the-envelope calculation shows that even with an average characterisation time of 3 seconds per system state, the time taken to reach $\mathcal{O}(10^4)$ characterisations is over 8 hours. Previous work by Haskey \emph{et al.} uses data mining techniques to to extract plasma fluctuations in time-series data, allowing one to identify different events using unsupervised classification. \cite{haskey2014clustering,haskey2015experiment}

Machine learning (ML) also offers an increased capability to perform correlation studies in high-dimensional spaces. Here, we present correlation studies performed using NSTX data, and describe the newly developed ML framework ERICSON (Experimental Resonant Instability Correlation Studies on NSTX), examining plasma wave frequency-chirping observed via Mirnov coils. ERICSON allows us to compare NSTX data with a wide range of parameters that would be largely unfeasible with human classification, allowing us to understand which parameters affect EP transport. After initial training of the algorithm, we find the time taken for a single process to reach $\mathcal{O}(10^4)$ characterisations is under 10 minutes, allowing for one to examine a much higher number of characterisations, and therefore a much richer set of correlations. While ML is not asymptotically convergent to perfect accuracy, it allows for broad, statistical recognition of the plasma stability boundaries that do exist. One expects any erroneous characterisations to still allow for asymptotically correct stability boundaries as one tends to an infinite number of characterisations.

\begin{figure}
	\begin{subfigure}{0.45\textwidth}
		\includegraphics[width=\textwidth]{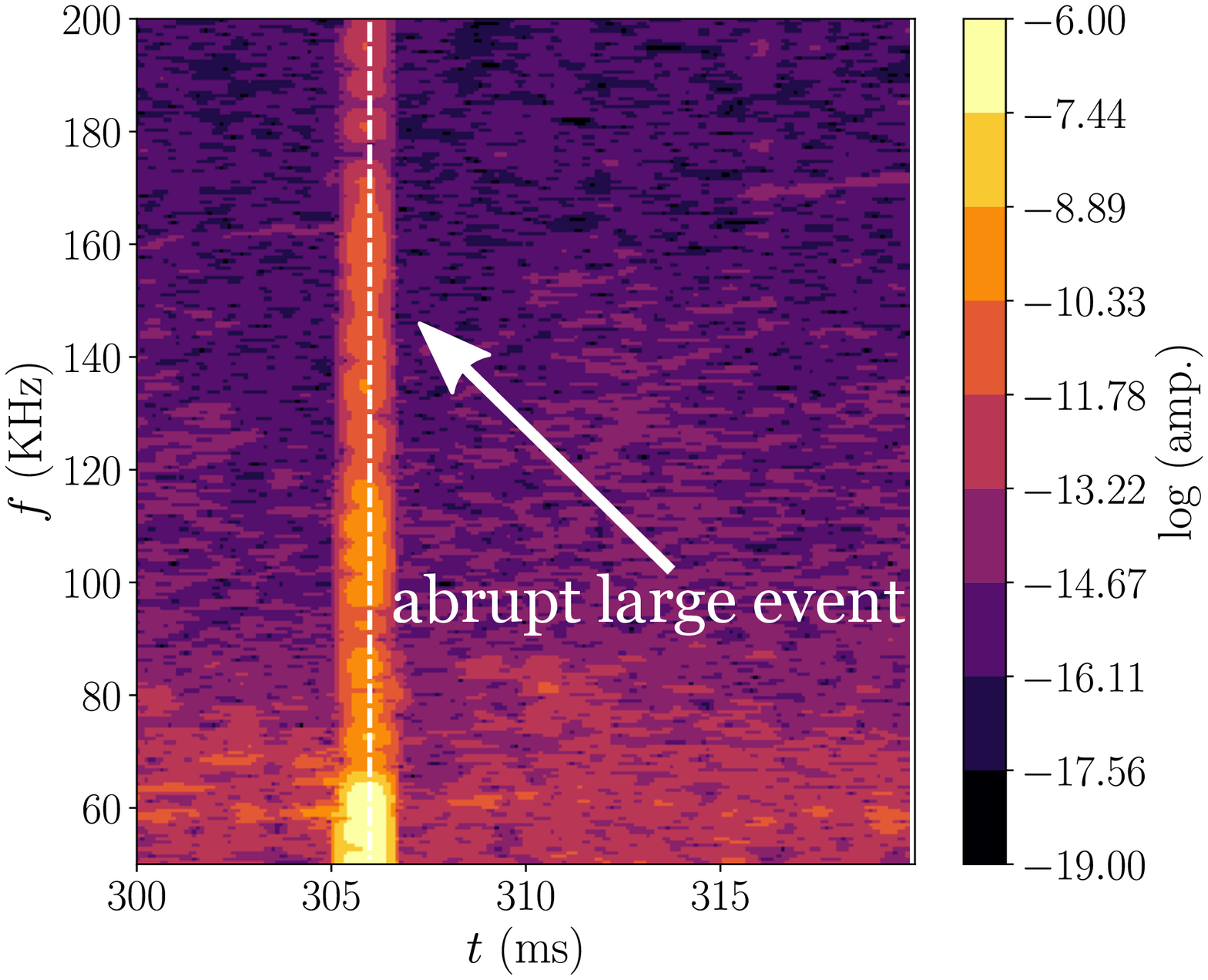}
	\end{subfigure}
	\begin{subfigure}{0.45\textwidth}
		\includegraphics[width=\textwidth]{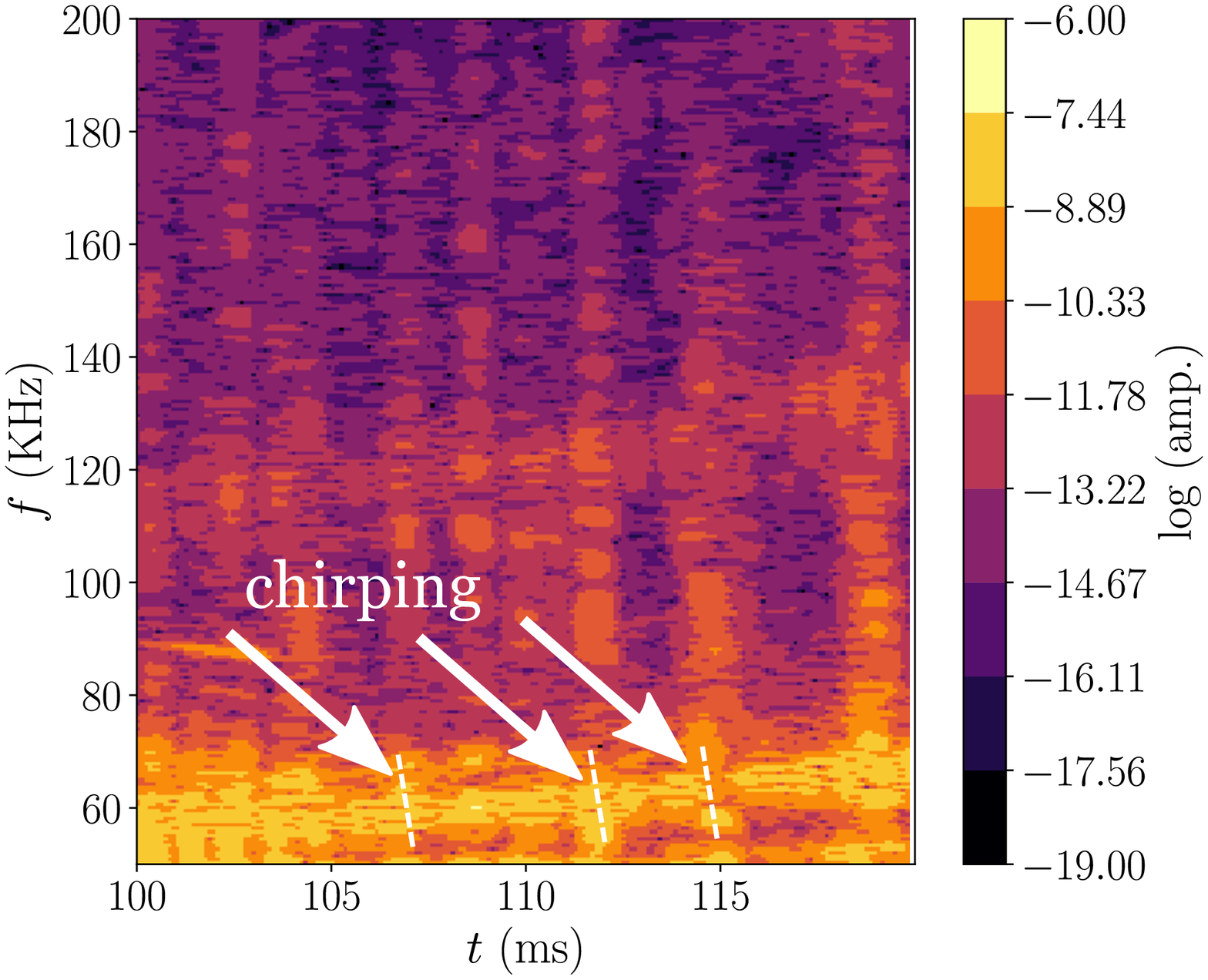}
	\end{subfigure} \\
	\begin{subfigure}{0.45\textwidth}
		\includegraphics[width=\textwidth]{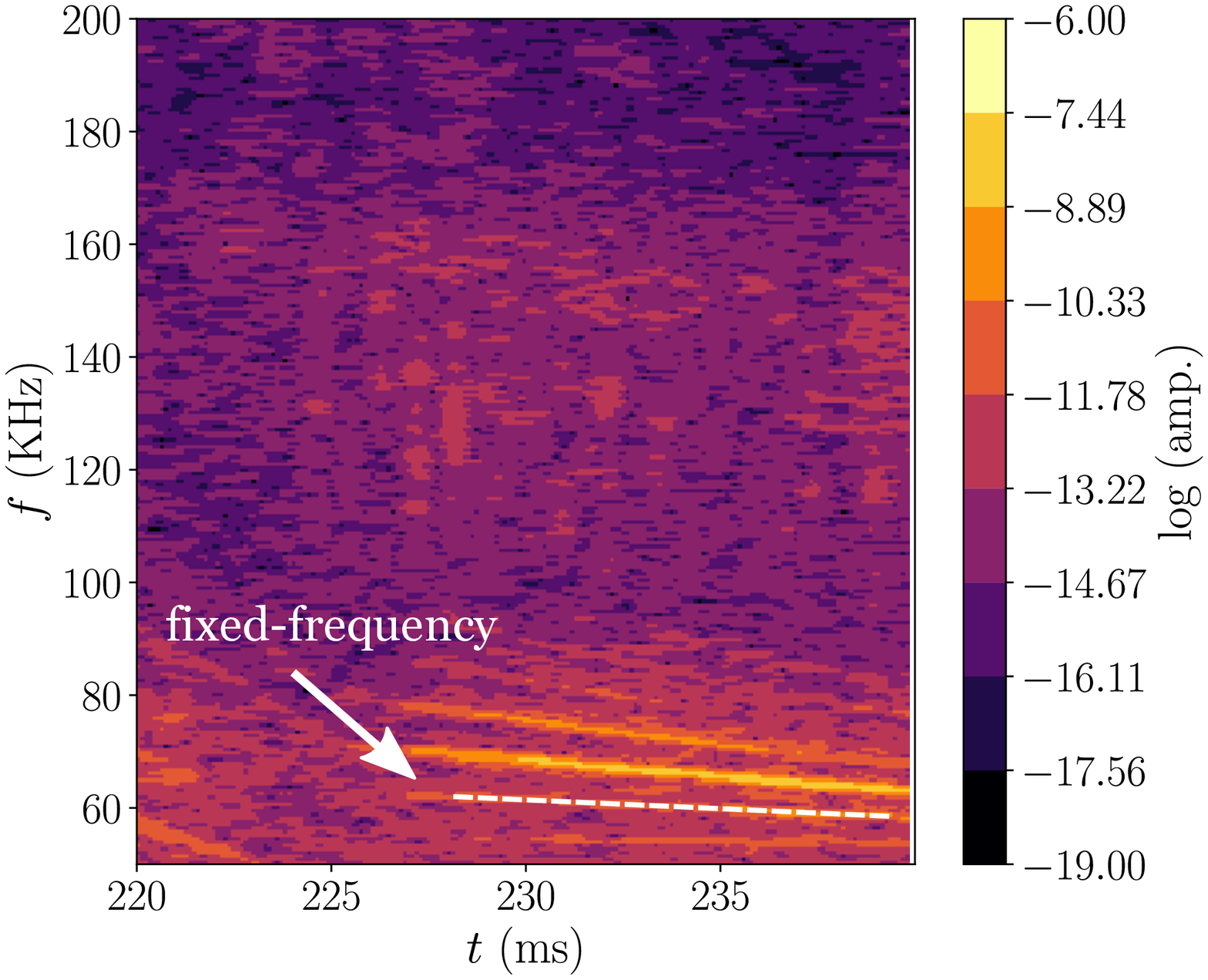}
	\end{subfigure}
	\begin{subfigure}{0.45\textwidth}
		\includegraphics[width=\textwidth]{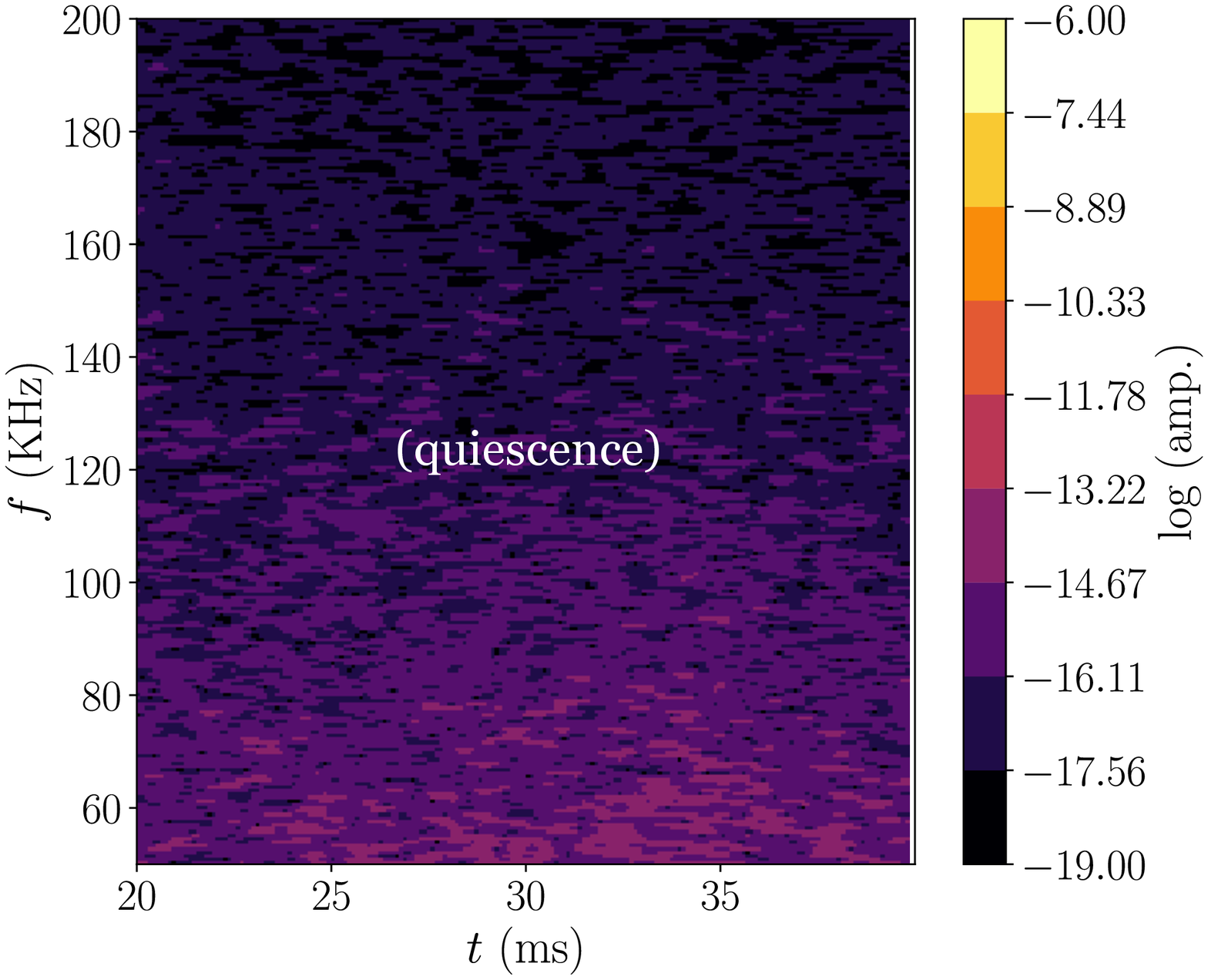}
	\end{subfigure}
\caption{\label{fig:char} Plots showing the 4 characterisations of toroidal Alfv\'{e}nic eigenmode (TAE) magnetic fluctuations obtained from shot 139317 on NSTX. Top left: abrupt large event, top right: chirping, bottom-left: fixed-frequency, bottom-right: quiescence.}
\end{figure}

Magnetic perturbations in a very broadband range (0 to $\sim$ \SI{500}{\mega\hertz}) are commonly measured on tokamaks such as NSTX by using Mirnov coils. Here, 5 characterisations of the frequency of magnetic perturbations are examined for KTF modes (of which 3 are shown in \figref{fig:KTF}): quiescence (noise, or no frequency dependent behaviour), approximately fixed-frequency eigenmodes (herein referred to as simply fixed-frequency), sweeping eigenmodes (slow frequency variation due to time evolution of the plasma equilibrium), chirping (rapid frequency variation over a narrow frequency band due to wave-particle interaction), and fishbone-like (energetic particle mode with rapid frequency variation over a broad frequency band). For TAEs, we also examine 4 characterisations (see \figref{fig:char}): quiescence, fixed-frequency, chirping, and ALEs (rapid frequency variation over a broad frequency band). Data from NSTX experiments in 2010 is utilised, revealing a rich set of correlations between different mode character and weighted averages of plasma parameters obtained from TRANSP \cite{code_transp}. Our results are in agreement with previous work by Fredrickson et al. \cite{fredrickson2014parametric} using human characterisation in a reduced parameter space, allowing us to heuristically confirm the validity of predictions made by ERICSON. We show new, strong correlations between moments of the spectrograms and mode character, as well as evidence of a TAE stability boundary along $v_{\varphi} \approx \frac{1}{4}(v_{\textrm{inj.}} - 3v_{\textrm{A}})$.

\section{\label{sec:theory} Fast ion loss}
Fast ions carry significantly more energy per mole than the thermal ions; on NSTX the NBI peak energy ($U_{\text{inj.}}$) lies at around \SI{90}{\kilo\electronvolt} and the thermal peak lies at $\sim$ \SIrange{1}{2}{\kilo\electronvolt} during typical tokamak operation \cite{synakowski2003national}. The overall plasma performance can be heavily reduced in cases of large fast ion losses.\cite{darrow2013stochastic} Fast ions give a significant contribution to the plasma pressure and are essential to transfer heat to the thermal population.

During frequency chirping events, non-linear structures known as `holes' and `clumps' can form on the ion distribution function, existing respectively as a relative decrease and increase of the local distribution function.\cite{berk1995numerical,berk1996nonlinear,berk1997spontaneous,todo2019introduction} As the hole and clump move along the distribution function, they drag the resonant wave frequency with them. The canonical momentum space drift leads to a kinetic transport process; as a particle decreases in $p_{\varphi}$, it moves to a flux surface at higher minor radius.

NBI heating and RF heating can lead to a quasi-steady `slowing-down' distribution of the form \cite{cordey1976effects,gaffey1976energetic}:

\begin{equation}
\label{eq:slowing}
F_{Si}(v) = \dfrac{3}{4 \pi \ln (1 + v_{\textrm{c}}^3/v_{\textrm{inj.}}^3)} \dfrac{n_{i}}{v_{\textrm{c}}^3 + v^3} H(v_{\textrm{inj.}} - v)
\end{equation}

where $F_{Si}$ is the slowing-down distribution function for the $i^{\textrm{th}}$ fast ion species, $H$ is the Heaviside step function, $n_{i}$ is the equilibrium number density of the $i^{\textrm{th}}$ fast ion species, and $v_{\textrm{c}}$ is the so-called critical speed:

\begin{equation}
v_{\textrm{c}} \equiv v_{\textrm{te}} \left( \dfrac{3 \sqrt{\pi}}{4} \sum\limits_i \dfrac{n_i m_{\textrm{e}}}{n_{\textrm{e}} m_i} Z_i^2 \right)^{1/3}
\end{equation}

where $v_{\textrm{te}}$ is the electron thermal speed, $n_{\textrm{e}}$ is the equilibrium electron number density, $m_{\textrm{i}}$ is the mass of the $i^{\textrm{th}}$ fast ion species, $m_{\textrm{e}}$ is the electron rest mass, and $Z_i$ is the atomic number of the $i^{\textrm{th}}$ fast ion species. Here, quasi-steady refers to the approximation that the distribution evolves smoothly as a function of $P_{\textrm{inj.}}$; fundamentally, this assumes that the rate of increase of injected power $1/\tau_{\textrm{ramp}}$ is much less than the reciprocal of the Spitzer slowing down time.

\clearpage

\onecolumngrid

\twocolumngrid

In the presence of background dissipation, kinetic instabilities that lead to hole and clump formation are triggered by gradients in the toroidal canonical angular momentum in a manner akin to inverse Landau damping; slowing down distributions feature large gradients in the neighbourhood of $v_{\textrm{inj.}} \equiv \sqrt{2 U_{\textrm{inj.}} / m}$, and therefore wave-particle interaction is enabled for waves with a resonance near $v_{\textrm{inj.}}$. As one attempts to increase the core temperature of the plasma, one creates sharper momentum gradients, leading to greater plasma instability. On NSTX, the NBI beam energy is super-Alfv\'{e}nic ($v_{\textrm{inj.}} > v_{\textrm{A}}$), which allows for increased kinetic instability of Alfv\'{e}n waves.

The formation of gap TAEs (gap toroidal Alfv\'{e}n eigenmodes) allow for relatively long lifetime waves; these waves are not dispersive in radius, and therefore allow for long range frequency chirping in the $\sim$ \SIrange{50}{200}{\kilo\hertz} range \cite{heidbrink2008basic}; the larger the frequency chirp, the further the momentum drift of the resonant particles. 

\begin{figure}
	\begin{subfigure}{0.45\textwidth}
		{\Large a)} \begin{tabular}{c|c}
		\centering
		KTF character & Frequency traces \\
		\hline
		{\color{green} $\bullet$} (quiescent) & Noise \\  
		{\color{cyan} $\bullet$} (fixed-frequency) & Constant $\omega(t)$ \\
		{\color{orange} $\bullet$} (sweeping) & Varying $\omega(t)$ \\ & ($>>$ \SI{1}{\milli\second}) \\
		{\color{blue} $\bullet$} (chirping) & Rapidly varying $\omega(t)$ \\ & ($\sim$ \SI{1}{\milli\second})\\
		{\color{magenta} $\bullet$} (fishbone-like) & Rapidly varying $\omega(t)$ \\ & (broadband, $\sim$ \SI{1}{\milli\second})
		\end{tabular}
	\end{subfigure} \\
		\vspace{15pt}
	\begin{subfigure}{0.45\textwidth}
		{\Large b)} \begin{tabular}{c|c}
		\centering
		TAE character & Frequency traces \\
		\hline
		{\color{green} $\bullet$} (quiescent) & Noise \\  
		{\color{cyan} $\bullet$} (fixed-frequency) & Constant $\omega(t)$ \\
		{\color{blue} $\bullet$} (chirping) & Rapidly varying $\omega(t)$ \\ & ($\sim$ \SI{1}{\milli\second}) \\
		{\color{magenta} $\bullet$} (ALEs) & Rapidly varying $\omega(t)$ \\ & (broadband, $\sim$ \SI{1}{\milli\second})
		\end{tabular}
	\end{subfigure}
    \caption{\label{fig:char_desc} a): Table detailing the characterisations of the mode frequency explored for kink/tearing/fishbones. The traces here refer to continuous streaks of slowly changing colour as one progresses in time on $\hat{B}(f,t)$ spectrograms. b): Table detailing the characterisations of the mode frequency explored for TAEs.}
\end{figure}

Furthermore, experiments and simulations in the literature \cite{fredrickson2006fast,darrow2013stochastic,kramer2017improving,collins2017phase} have shown that rapid long range frequency chirping across a wide range of frequency chirps (mode `avalanching') is correlated with high amounts of fast ion loss. Other work has shown that mode-mode destabilisation may play a role in Alfv\'{e}nic frequency chirping - activity in the KTF frequency range ($\sim$ \SIrange{1}{30}{\kilo\hertz}) may instigate Alfv\'{e}nic activity, and vice-versa.

Crucially, the conditions which trigger small scale frequency chirping and so-called mode avalanching are not fully understood. Intuitive knowledge related to momentum gradients provides insight, but factors such as the $q$-profile, energy confinement time, and the fast ion density gradient may also play a role.

\section{\label{sec:framework} Framework}
ERICSON utilises four key parts: pre-processing NSTX data, mode characterisation, parameter space tracking from TRANSP data, correlation studies. Here, we discuss the pre-processing, mode characterisation, and the parameter space tracking; later, we show results from some correlation studies.

\subsection{Pre-processing (NSTX data)}
The induced voltage was measured across Mirnov coils. Via the Faraday-Lenz law:

\begin{equation}
V \equiv \oint\limits_{\partial A} \mathbf{E} \cdot \dif \boldsymbol{\ell} = - \partial_t \int\limits_{A} \mathbf{B} \cdot \dif\mathbf{A}
\end{equation}

where $V$ is the induced voltage, $\partial A$ is the boundary of the area $A$ of the coil, $\mathbf{E}$ is the electric field, $\mathbf{B}$ is the magnetic flux density, and $\hat{\mathbf{a}}$ is the unit vector normal to the surface, such that $\dif \mathbf{A} = \hat{\mathbf{a}} \dif A$.

We utilise the following asymmetric definitions for the discrete Fourier transform:

\begin{align}
\mathcal{F}[B](\mathbf{r}, f[j]) &= \sum\limits_{l = 0}^{N_t - 1} B(\mathbf{r},t[j]) \exp \left(\rmi 2 \pi f[j] t[l]\right) \\
\mathcal{F}^{-1}[\tilde{B}](\mathbf{r}, t[j]) &= \dfrac{1}{L_t} \sum\limits_{l = 0}^{N_t - 1} \tilde{B}(\mathbf{r},f[l]) \exp \left(-\rmi 2 \pi f[l] t[j] \right)
\end{align}

where $\mathbf{r}$ is the position in real space, $f = \{l \in Z: 0 \leq l < N_t \}$ is a set of $N_t$ frequencies, $N_t$ is the number of time points in the dataset, and $L_t$ is the temporal length of the dataset. Then, Fourier transforming the Faraday-Lenz law yields:

\begin{equation}
\tilde{V} \propto - \rmi 2 \pi f \tilde{B}
\end{equation}

One can immediately see that $\tilde{B}$ is singular at $f = 0$. Because discrete analysis is used, this singularity becomes broadened and can affect nearby points. For this reason, we employ a cut off frequency ($f_{\textrm{min.}} \equiv$ \SI{1}{\kilo\hertz}) to avoid the singular value and nearby points saturating the data set. To minimise pick-up from the error-field correction switching power amplifiers (SPAs), we take the average time-domain signal from two Mirnov coils in close proximity of each other. However, one may instead choose to produce spectrograms from the cross-correlation $S_{\textrm{cr.}}$ of the two signals $\{V_1, V_2\}$ such that:

\begin{equation}
S_{\textrm{cr.}}(t) = \int\limits_{-\infty}^{\infty} V_1(\tau) V_2(\tau + t) \dif \tau
\end{equation}

Under Fourier transform, this yields the product of the Fourier transform of the two signals. Here, we assume that the variation in signal amplitude and phase is small between the two coils, such that:

\begin{equation}
|\tilde{V}(\omega)| = \left|\dfrac{\tilde{V}_1 + \tilde{V}_2}{2}\right| \approx \sqrt{|\tilde{S}_{\textrm{cr.}}(\omega)|}
\end{equation}

As a result, for cases where the signal amplitude or phase varies appreciably between the two coils, it may prove advantageous to use the cross-correlation instead (for further details, please see \appref{app:cr}). To enable analysis of time-dependent frequencies, we use a short-time Fourier transform (STFT) to track the frequency evolution of modes in the system. The frequency resolution $\Delta f$ and maximum frequency $f_{\textrm{max.}}$ obtained via an STFT are given respectively by:

\begin{align}
\label{eq:omega}
\Delta f = \dfrac{1}{L_t} &;& f_{\textrm{max.}} = \dfrac{1}{2\Delta t}
\end{align}

where $\Delta t$ is the time resolution of the data set, and the data set is now more specifically defined as the time points within the STFT window. The forward-difference STFT is calculated as follows:

\begin{equation}
\hat{V}(f, t) = \mathcal{F}\bigg[\big\{W(t') V(t') : t < t' < t + L_t\big\}\bigg]
\end{equation}

where $t$ is the time at which the STFT begins, $t'[j] = j \Delta t'$ is a dummy time used for the transform, $W(t')$ is a window function, and $V(t)$ is now the two Mirnov coil average of the induced voltage. Many window functions are employed in the literature \cite{harris1978use}, each producing spurious sidebanding for a single frequency input. Here we employ a Hanning window due to its favourable decibel tapering for the erroneous signal produced, and its simple form:

\begin{equation}
W(t') = \sin^2 \left( \dfrac{\pi t'}{N_t - 1}\right)
\end{equation}\\

Using this method, one obtains the spectrogram signal:

\begin{equation}
\label{eq:signal}
\hat{B}(f, t) \propto \dfrac{\rmi}{2 \pi f} \hat{V}(f, t)
\end{equation}

Finally, one can separate the spectrogram into different subdomains of $f$, allowing for characterisation of the signal in different frequency bands. A human would typically examine spectrograms obtained from this pre-processing, and perform characterisation (see \figref{fig:char} and \figref{fig:char_desc} for characterisations).

For ML training and analysis, 125 shots from the 2010 NSTX archives were selected due to the clear observation of KTF and TAE activity. The Mirnov coils produce a signal at sampling rate $f_{V} =$ \SI{5}{\mega\hertz} (\SI{200}{\nano\second} resolution). 

The shots were split into \SI{20}{\milli\second} slices, allowing for a maximum of $\sim 20$ chirping events; typically chirping in these frequency bands occurs on a $\sim$ \SI{1}{\milli\second} timescale. 

The STFT time window $L_t$ contained $2^{13}$ samples (\SI{1.6384}{\milli\second}), allowing for a frequency resolution of $\sim$ \SI{0.610}{\kilo\hertz}. By sliding the STFT using increments of $2^9$ samples, we were able to obtain a time resolution $\tau$ on the spectrogram of \SI{102.4} {\micro\second}. It is instructive to note that by using a sliding increment smaller than the STFT window, to produce overlaid spectrograms that are plotted here (such as in \figref{fig:char}) one must add a `lag' of \SI{819.2}{\nano\second} as the pixel is actually the pixel at the center of the STFT window.

The 2D data for each slice was then converted into a contiguous 1D array in memory, made by taking contiguous blocks in frequency of the original spectrogram. That is to say:

\[
\begin{array}{r l}
\textrm{ad.}[\hat{B}(f_2,t)] >> \textrm{sizeof}(\hat{B}) &= \textrm{ad.}[\hat{B}(f_1,t+\tau)] \\
\forall f < f_2: \textrm{ad.}[\hat{B}(f,t)] >> \textrm{sizeof}(\hat{B}) &= \textrm{ad.}[\hat{B}(f + \Delta f,t)]  
\end{array}
\]

where for a given spectrogram, $f_1$ is the minimum frequency, $f_2$ is the maximum frequency, $\textrm{ad.}[\hat{B}(f,t)]$ is the address in memory of the spectrogram signal at $(f,t)$, and $>>$ is the incremental bit-shift operator. A bit-shift by the datasize of the signal ($\textrm{sizeof}(\hat{B})$) allows us to relate two neighbouring array elements.

In total, a database of 4773 slices was generated. It is worth noting that there is a finite probability of an exact overlap of the plasma parameters between two slices in the database, although this is quite unlikely.

\subsection{ML algorithm and training}
In this paper, we employ supervised machine learning to characterise the spectral data. We have used random forest classifiers (RFCs), an ensemble variation of the well-established decision tree classifier (DTC). RFCs have been utilised and are well described in the literature \cite{breiman2001random, rea2018exploratory} allowing for a simple, white-box approach of the problem, something we believe to be important for correlation studies.

DTCs act as flowchart-like algorithms which are optimised using a greedy algorithm. Therefore, while these classifiers are easy to operate, they sometimes only find local minima in the Gini impurity, leading to erroneous categorisation. Increasing the complexity of the tree (adding depth to the decision tree) can sometimes `kick' the algorithm from a local minima into a global minima and improve accuracy, but too much depth leads to overfitting.

For this reason, we use RFCs which initialise an ensemble of decision trees with different random initial states. Then, the RFC returns the average probability that the data is of a given classification. This leads to a $\mathcal{O}(N_{\textrm{DTC}})$ increase in accuracy for small $N_{\textrm{DTC}}$, where $N_{\textrm{DTC}}$ is the number of trees.

Here, we examine a very high dimensional feature space (each spectrogram pixel is a single dimension) with an extremely high amount of collinearity; mode character is observed by relating the spectrogram amplitude of local clusters of pixels. RFCs subdivide the feature space on a Cartesian grid, with hyperparameters determining the available number of subdivisions (tree depth) and the size of the minimum undivided volume (leaf size).

Unfortunately, highly collinear, high dimensional spaces are hard for RFCs to efficiently divide. Therefore we expect RFCs to yield a lower accuracy than other ML schemes such as convolutional neural networks (CNNs). However for correlation studies, we examine broad, collective behaviour in parameter space; as a result we do not require extremely high accuracies.

The RFCs here are implemented using the Python library \verb|scikit-learn|\cite{pedregosa2011scikit}; it is feasible that one could perform similar analysis in \verb|tensorflow|, or write an RFC from scratch.

Human classification was performed using 10 shots, producing 337 slices for the KTF band (\SIrange{1}{30}{\kilo\hertz}) and 281 slices for the TAE band (\SIrange{50}{200}{\kilo\hertz}). We utilised training slices which featured low ambiguity in the frequency behaviour; some of these slices included multiple characterisations during the same time slice.

\begin{figure}[h]
	\centering
	\includegraphics[width=0.45\textwidth]{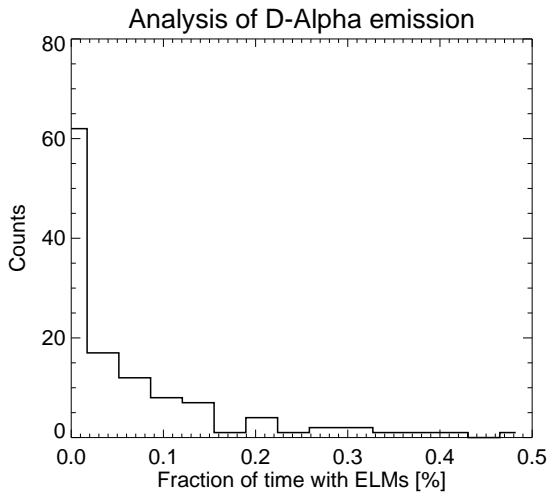}
	\caption{Histogram for the number of shots with ELMs (edge localized modes) versus the fraction of time points in that shot with ELMs, using spikes in the $D_{\alpha}$ emission obtained from a fast ion $D_{\alpha}$ (FIDA) diagnostic. The shots featured here are the shots used in the correlation analysis performed in \secref{sec:corr} after training ERICSON. The fraction of time points containing ELMs is generally much less than 1\% for the shots examined in this work, and as such we assume that any misclassification of ELMs as abrupt large events can be considered a negligible contribution to the correlations observed in \secref{sec:corr}.}
	\label{fig:ELM}
\end{figure}

It is worth noting that here ERICSON does not check the $D_{\alpha}$ signal (light emission detected from the Balmer-alpha line) against the classification. As such, ERICSON could in principle classify ELMy time slices as featuring ALEs. However, the fraction of time points with ELMs, $F$, is very small. In \figref{fig:ELM}, we present $F$ using a histogram given by binning across different shots:

\begin{equation}
F = \dfrac{n_{\textrm{ELM}}}{f_D(t_1-t_0)}
\end{equation}

where $n_{\textrm{ELM}}$ is the number of time points with an ELM spike, $f_{D}$ is the sampling frequency of the fast ion $D_{\alpha}$ diagnostic, and $t_1-t_0$ is length of the time window of interest for each shot. In general, $F \ll 1\%$, and therefore we find that for the overall correlations given in \secref{sec:corr}, checking against the $D_{\alpha}$ signal was not deemed necessary for this work. However, for ELMy discharges, one would need to consider the $D_{\alpha}$ signal. We desire that characterisation falls under the following hierarchy for the KTF band:

\[
\textrm{fishbone-like} \to \textrm{chirping} \to \textrm{sweeping} \to \textrm{fixed-freq.} \to \textrm{quie.}
\]

with leftmost as the most important feature. That is to say, we desire a multi-class classifier such that any ambiguity leads to a more leftmost characterisation. For the TAE band, we desire the following hierarchy:

\[
\textrm{ALEs} \to \textrm{chirping} \to \textrm{fixed-freq.} \to \textrm{quie.}
\] 

with leftmost as the most important feature. A set of RFCs were then trained to perform binary classification separately; each RFC yields a binary output for whether the system is exhibiting each of the types of character. We employ separate binary classification to allow us to tweak multi-class classification in a more direct fashion. The RFCs were trained using 75\% of the learning set, and tested on the remaining 25\%. The tree depth and number of trees were manually tweaked to optimise multi-class classification accuracy.

\begin{figure}[h]
	\centering
	\includegraphics[width=0.45\textwidth]{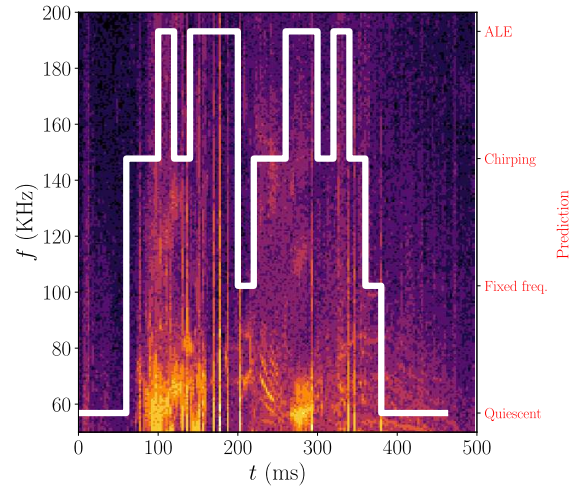}
	\caption{Frequency spectrogram for magnetic fluctuations on NSTX in the \SIrange{50}{200}{\kilo\hertz} (toroidal Alfv\'{e}n eigenmode) band for the first 500ms of shot 139317, overlaid with the mode character classified by ERICSON. 4 types of mode character are observed: quiescence, fixed-frequency eigenmodes, chirping modes, and abrupt large events (ALEs).}
	\label{fig:overlay}
\end{figure}

In \figref{fig:overlay}, we show a frequency spectrogram for magnetic fluctuations found during NSTX shot 139317, overlaid with the classification made by ERICSON. As can be seen, ERICSON categorizes fairly cautiously, and sometimes characterises chirping behaviour as ALEs. To classify, a na\"{i}ve approach might simply rely on using the most likely classification. In pseudocode one can write this as:

{\small
\begin{lstlisting}
for (i = 0; i < K; i ++) {
  prob[i] = RFC(i)
}
return max_index(prob)
\end{lstlisting}}

\clearpage

\onecolumngrid

\begin{figure}
	\begin{subfigure}{0.45\textwidth}
	$\begin{array}{r c l}
	C_{\text{KTF}}^{(\text{train})}: \left[\begin{matrix}
	140 & 0 & 0 & 0 & 0 \\
	1 & 54 & 0 & 0 & 0 \\
	0 & 0 & 12 & 0 & 0 \\
	0 & 0 & 0 & 10 & 0 \\
	0 & 0 & 0 & 0 & 18
	\end{matrix}\right] &;&
	C_{\text{KTF}}^{(\text{test})}: \left[\begin{matrix}
	57 & 2 & 1 & 0 & 4 \\
	1 & 16 & 3 & 1 & 0 \\
	2 & 0 & 2 & 0 & 2 \\
	0 & 0 & 0 & 1 & 1 \\
	2 & 0 & 0 & 0 & 8
	\end{matrix}\right]
	\end{array}$
	\end{subfigure} 
	\hspace{20pt}
	\begin{subfigure}{0.45\textwidth}
	$\begin{array}{r c l}
	C_{\text{TAE}}^{(\text{train})}: \left[\begin{matrix}
	138 & 0 & 0 & 0 \\
	0 & 12 & 0 & 0 \\
	0 & 1 & 22 & 0 \\
	0 & 0 & 0 & 24
	\end{matrix}\right] &;&
	C_{\text{TAE}}^{(\text{test})}: \left[\begin{matrix}
	59 & 0 & 0 & 1 \\
	1 & 2 & 1 & 2 \\
	0 & 2 & 7 & 1 \\
	0 & 0 & 3 & 6 
	\end{matrix}\right]
	\end{array}$
	\end{subfigure}\\
	\vspace{10pt}
	\begin{subfigure}{0.45\textwidth}
	\begin{tabular}{c|c|c}
	Metric & Training set & Test set \\
	\hline
	Accuracy & 99.2\% & 69.0\% \\
	$R_K$ & 0.996 &0.824
	\end{tabular}
	\end{subfigure}
	\hspace{20pt}
	\begin{subfigure}{0.45\textwidth}
	\begin{tabular}{c|c|c}
	Metric & Training set & Test set \\
	\hline
	Accuracy & 98.9\% & 72.6\% \\
	$R_K$ & 0.995 &0.871
	\end{tabular}
	\end{subfigure}
\caption{Confusion matrices and useful metrics from ERICSON trained to predict different types of classification for magnetic fluctuations on NSTX in the \SIrange{1}{30}{\kilo\hertz} (kink/tearing/fishbone) band and \SIrange{50}{200}{\kilo\hertz} (toroidal Alfv\'{e}n eigenmode) band. From top-left to bottom-right, the leading diagonal of the confusion matrix for KTFs corresponds to correct classification for quiescence, fixed-frequency, frequency sweeping, chirping, and fishbone-like. From top-left to bottom-right, the leading diagonal of the confusion matrix for TAEs corresponds to correct classification for quiescence, fixed-frequency, chirping, and abrupt large events. The confusion matrices have high sparsity in the lower-diagonal part, owing to the preferential bias incorporated in ERICSON to increase prediction accuracy and enforce safer predictions.}
\label{fig:confusion}
\end{figure}

\twocolumngrid

where $K$ is the number of binary classifiers used. This is typically the method employed by most multi-class classifiers. The classification accuracy from this method is typically $\textrm{min}(\{p_i\})$ at best where $p_i$ is the accuracy of the $i^{\textrm{th}}$ RFC, making it a reasonable approach. However, due to the desired hierarchy, we instead impose a heirarchal method:

{\small
\begin{lstlisting}
for (i = 0; i < K; i++) {
  if RFC(i) > RFC(l) + tol(i) for all l > i {
    return i
  }
}
\end{lstlisting}}

Here, \verb|tol(i)| denotes a tolerance factor. The confusion matrix $\{C_{ij}\}$ is defined here as the number of predictions of class $j$ that were actually of class $i$, and the tolerance factor for each characterisation is therefore prescribed such that the confusion matrices for the training and test set meet two constraints: that the confusion matrices are close to upper-triangular, and that the confusion matrices are as close to diagonal as possible. The latter is the most important constraint, as a diagonal confusion matrix denotes perfect characterisation. By using the tolerance factor, we can force ERICSON to make better decisions by making it `more cautious'; ERICSON would then preferentially classify TAE activity in a slice as chirping when previously classified as quiescent, if the tolerance is set to be higher. For moderate tolerance levels, this reduces overall confusion between different characterisations, but naturally creates some off-diagonal elements in the above diagonal part of the confusion matrices. This is deemed to be an acceptable error as we have encoded a preference for caution; we would rather have ERICSON predict a more `dangerous' behaviour if it does misclassify.

In general, many of the characterised slices will have multiple features. Unfortunately this means that while the `most likely' method produces the best accuracy out of the multi-class algorithms we employed, it has a finite ceiling on the accuracy, owed to the fact that it will not adequately distinguish between two features which appear at the same time during a slice. While the ML multi-class classifier could be improved, we find the accuracy to still be acceptable. In kind, a human performing classification could quite easily misinterpret a slice; what may be recorded as slow frequency sweeping by one human could be recorded as fixed-frequency by another. For this reason, we include in \figref{fig:confusion} the more conventional metric of accuracy, as well as the $R_K$ coefficient. This coefficient has an upper limit of 1, and a lower limit which is greater than or equal to -1 depending on the dataset \cite{gorodkin2004comparing}:

\begin{widetext}
\begin{equation}
R_K = \dfrac{\sum_{klm} C_{kk} C_{lm} - C_{kl} C_{mk}}{\sqrt{\sum_k (\sum_l C_{kl}) \left(\sum_{l', k' = k} C_{k' l'}\right)} \sqrt{\sum_k (\sum_l C_{lk}) \left(\sum_{l', k' = k} C_{l' k'}\right)}}
\end{equation}
\end{widetext}

The $R_K$ coefficient is a generalisation of the Matthews Correlation Coefficient for multi-class classification. It is intrinsically related to the Pearson correlation coefficient, as it is directly expressed as the ratio of the covariance of two $K$-dimensional data sets corresponding to predicted classifications $(\underline{\mathbf{X}})$ and actual classifications $(\underline{\mathbf{Y}})$, and the product of the standard deviations of those two data sets:

\begin{equation}
R_K = \dfrac{\textrm{cov}(\underline{\mathbf{X}},\underline{\mathbf{Y}})}{\sqrt{\textrm{cov}(\underline{\mathbf{X}},\underline{\mathbf{X}}) \cdot \textrm{cov}(\underline{\mathbf{Y}},\underline{\mathbf{Y}})}}
\end{equation}

where $\textrm{cov}(a,b)$ is the covariance of $a$ and $b$. Because an unbalanced data set is used for training, we incur a reasonable amount of accuracy bias; the real data set is also unbalanced (for example, ALEs are much less common than periods of quiescence). While the $R_K$ is not as intuitive as accuracy, this is a better measure of the quality of the ML classifier, as it weights the metric depending on this accuracy bias, and gives a metric whose value is resilient under the use of an unbalanced real data set.

\subsection{Parameter space tracking (TRANSP data)}
For each discharge, plasma parameters are obtained from time-dependent simulations through the tokamak transport code TRANSP \cite{hawryluk1981empirical,code_transp}.

Simulations use measured profiles of electron and ion density and temperature, toroidal rotation, impurity content (assuming carbon is the dominant impurity). \cite{leblanc2003operation,bell2010comparison}

The plasma equilibrium as a function of time is computed through the EFIT code \cite{lao1985reconstruction} using measured coils and plasma currents, data from magnetic sensors and kinetic profiles as main constraints. When available, the q-profile measured using the motional Stark effect diagnostic (MSE) at the mid plane \cite{levinton2008motional} is also used as additional constraint.

Simulations include modeling of the fast ion distribution evolution in time through the Monte Carlo NUBEAM module of TRANSP \cite{goldston1981new,pankin2004tokamak}. NUBEAM includes classical phenomena such as collisional Coulomb scattering and slowing-down, and atomic physics (e.g. charge-exchange and neutralization reactions). Input parameters such as neutral beam injection energy, injected current, and neutral beam waveforms are taken from the experiment.

The main sources of uncertainty in TRANSP simulations are experimental uncertainties in the measured plasma profiles and inaccuracy in the equilibrium reconstruction. Typical uncertainties in the measured thermal plasma profiles are within $10 \%$ or better, which would result in small changes of derived quantities such as Alfv\'{e}n velocity and $\beta$ values reported in this work.

Similarly, uncertainties in the equilibrium reconstruction would cause a scatter of a few $\%$ in the reported values. For the equilibrium, the main cause of uncertainty is the lack of constraints on the q-profile in the core when MSE data are not available.

As one performs more classification, one populates the parameter space with more points. One expects that for phase transitions occurring over small, finite regions of parameter space, ERICSON should reveal regions with largely the same classification. Previous work has shown that frequency chirps exist for stochastic lifetimes \cite{woods2018stochastic}, with other work showing that stochastic transport mechanisms such as ion microturbulence can effect the likelihood of chirping and therefore the characterisations observed \cite{duarte2017prediction}.

For these reasons, we look at the overall behaviour of KTF and TAE magnetic activity in different regions of parameter space; one should not fully trust each individual classification, but rather the overall behaviour and the location of characterisation boundaries.

Each pixel shown in the figures in \secref{sec:corr} is binned such that the value of the pixel is the most frequent characterisation in that bin.

\section{\label{sec:corr} Correlations} 

\subsection{Mode-weighted averaging}
Chirping is a non-linear phenomenon, requiring wave-particle nonlinearity. If one aims to correlate chirping with plasma parameters, some of these parameters may be spatially dependent; accordingly the mode structure plays an important role.

Here, we model the spatial distribution of physical quantities dependent on Alfv\'{e}nic activity in the system via Bayesian inference, such that physical quantities are conditionally distributed \emph{given the mode structure}. As such, by modelling the mode structure by using a normalized Gaussian (and therefore taking the prior to be Gaussian), quantities dependent on Alfv\'{e}nic activity can be determined via a posterior distribution which is also Gaussian.

Accordingly, one can construct conditional expectations of the plasma quantities, where integrals are weighted with a normalised Gaussian, given here by $w(\rho)$:

\begin{equation}
\langle g \rangle \equiv \int\limits_0^1 (g \cdot w) \dif \rho
\end{equation}

where $\rho = \sqrt{\Psi/\Psi_0}$, $\Psi_0$ is the magnetic flux at the last closed flux surface, $g$ is the quantity to be spatially averaged, and $\langle g \rangle$ is the mode-weighted average.

Experimentally, the mode structure can be observed via reflectometry data, however this data cannot probe hollow density profiles. Unfortunately, for a sizable number of chirping cases, beam deposition and other effects can lead to hollow density profiles, preventing inference of the mode structure via reflectometry.

In lieu of fully reliable values for every shot analysed, we make three assumptions. First, one expects Alfv\'{e}n waves to have a fairly narrow mode structure, and as such we approximate the standard deviation to be $\sim 0.1$.

\clearpage

\onecolumngrid

\begin{figure}
	\begin{subfigure}[t]{0.47\textwidth}
	\includegraphics[width=\textwidth]{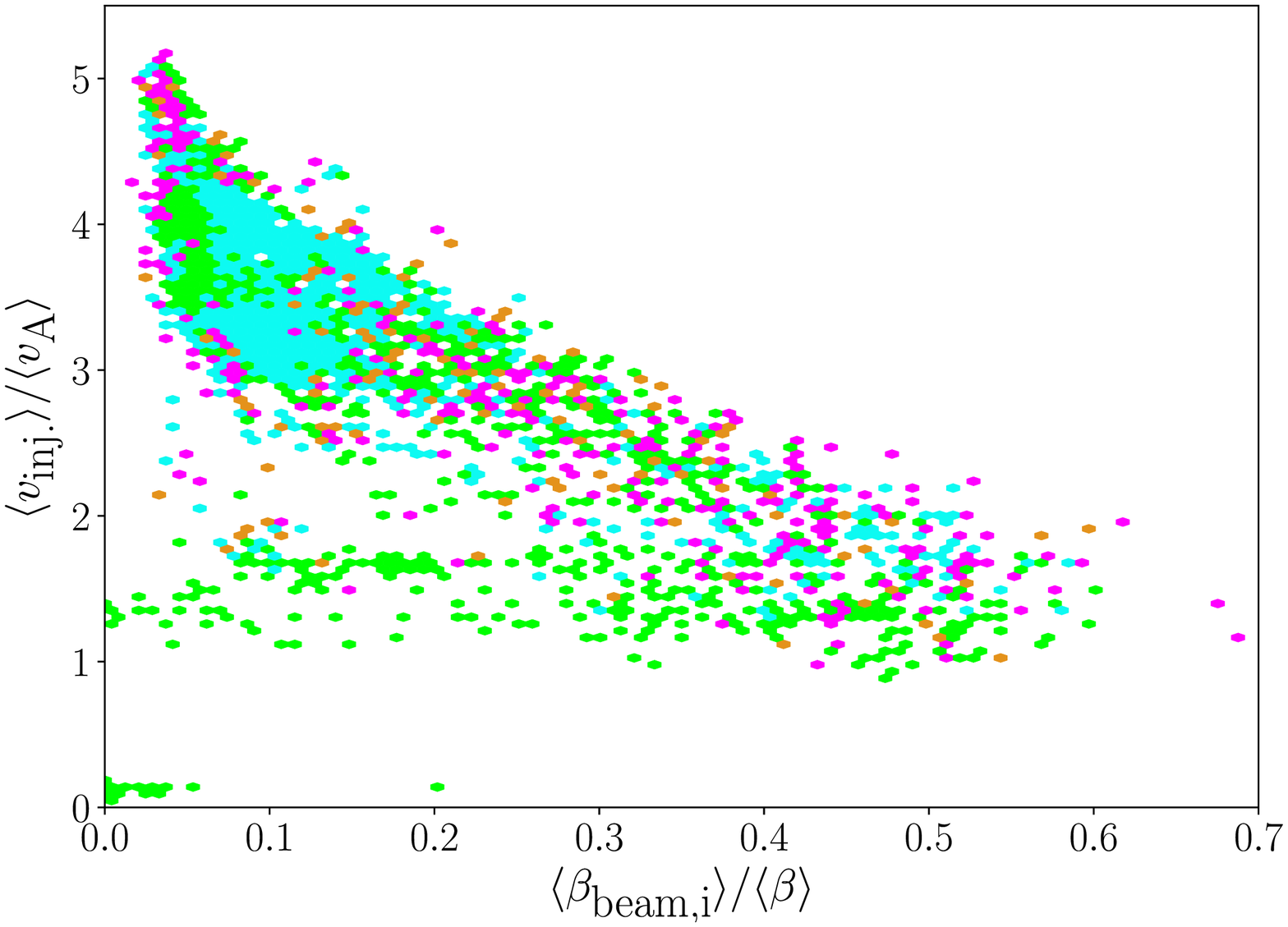}
	\caption{Normalized injection velocity versus normalized beam ion beta for kink/tearing/fishbones. Fixed-frequency modes largely confined to super-Alfv\'{e}nic injection velocity ($v_{\textrm{inj.}} \gtrapprox 2 v_{\textrm{A}}$) and low beam ion beta ($\beta_{\textrm{beam},i} \lessapprox 0.25 \beta$).}
	\end{subfigure}
	\begin{subfigure}[t]{0.47\textwidth}
	\includegraphics[width=\textwidth]{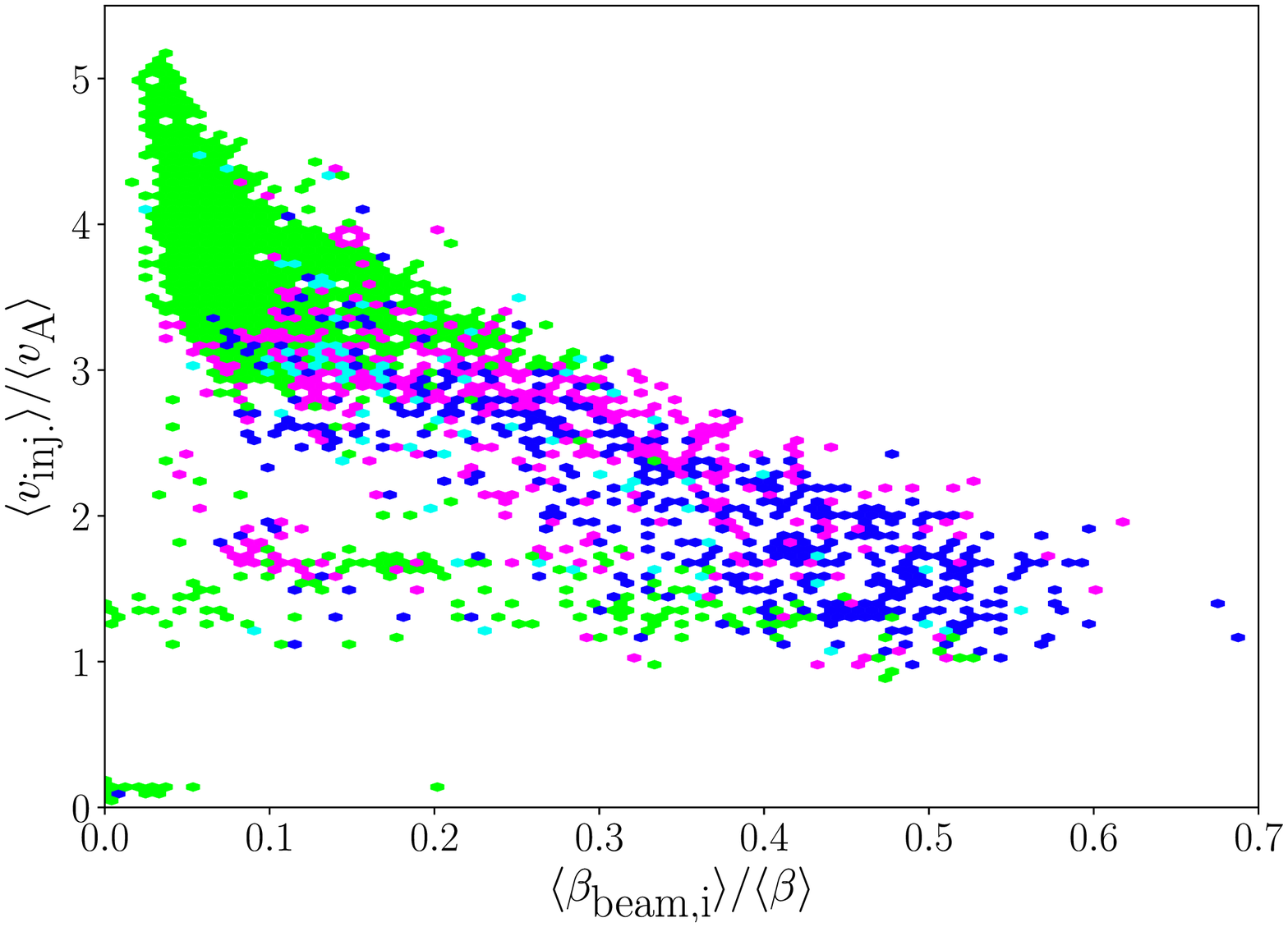}
	\caption{Normalized injection velocity versus normalized beam ion beta for TAEs. Quiescent behaviour largely confined to low beam ion beta ($\beta_{\textrm{beam},i} \lessapprox 0.2 \beta$).}
	\end{subfigure} \\
	\begin{subfigure}[t]{0.47\textwidth}
	\includegraphics[width=\textwidth]{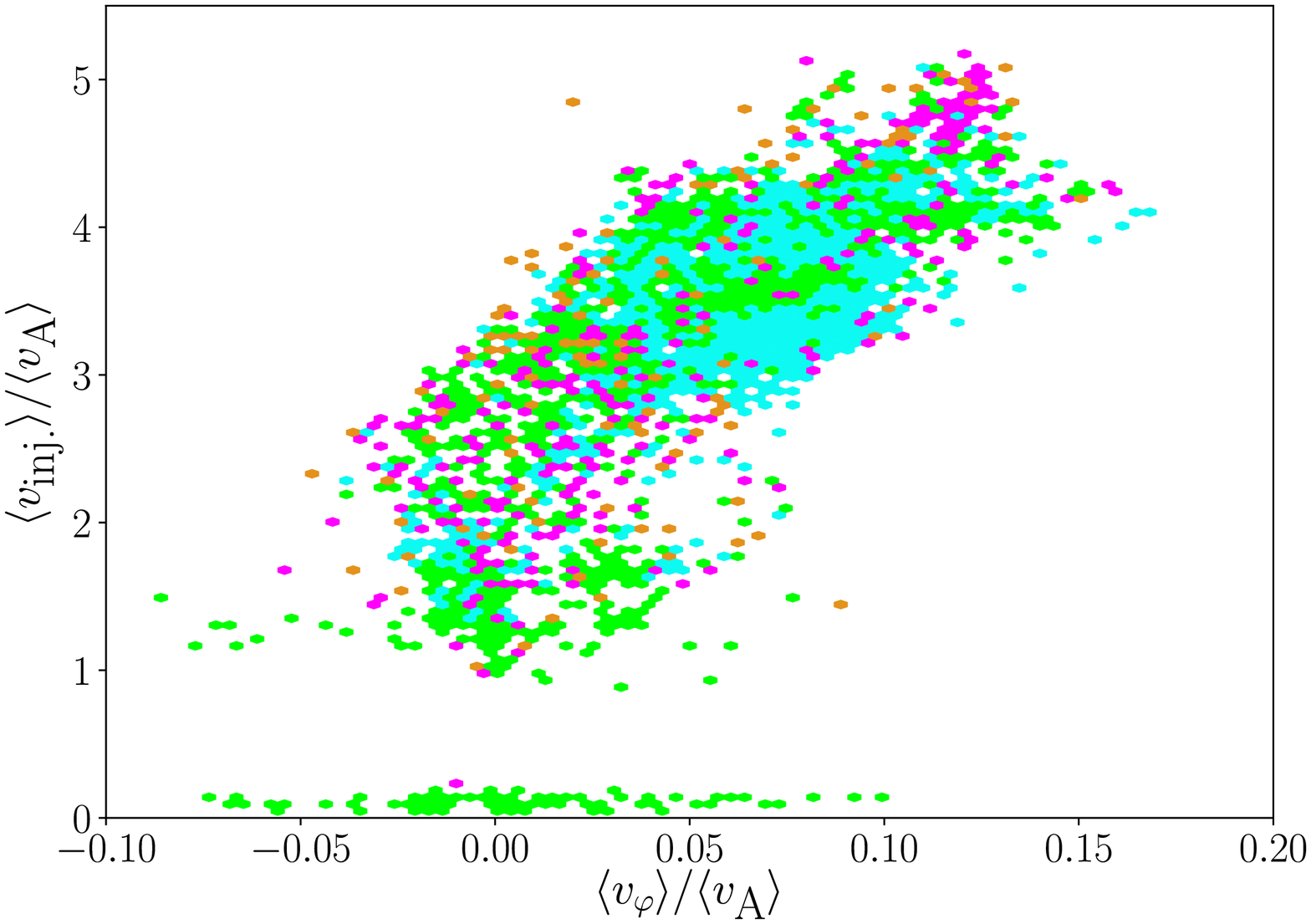}
	\caption{Normalized injection velocity versus normalized  plasma toroidal velocity for kink/tearing/fishbones. Quiescent behaviour largely confined to sub-Alfv\'{e}nic toroidal velocity ($v_{\varphi} \lessapprox 0.05 v_{\textrm{A}}$). Fixed frequency behaviour dominates for $v_{\varphi} \gtrapprox 0.05 v_{\textrm{A}}.$}
	\end{subfigure}
	\begin{subfigure}[t]{0.47\textwidth}
	\includegraphics[width=\textwidth]{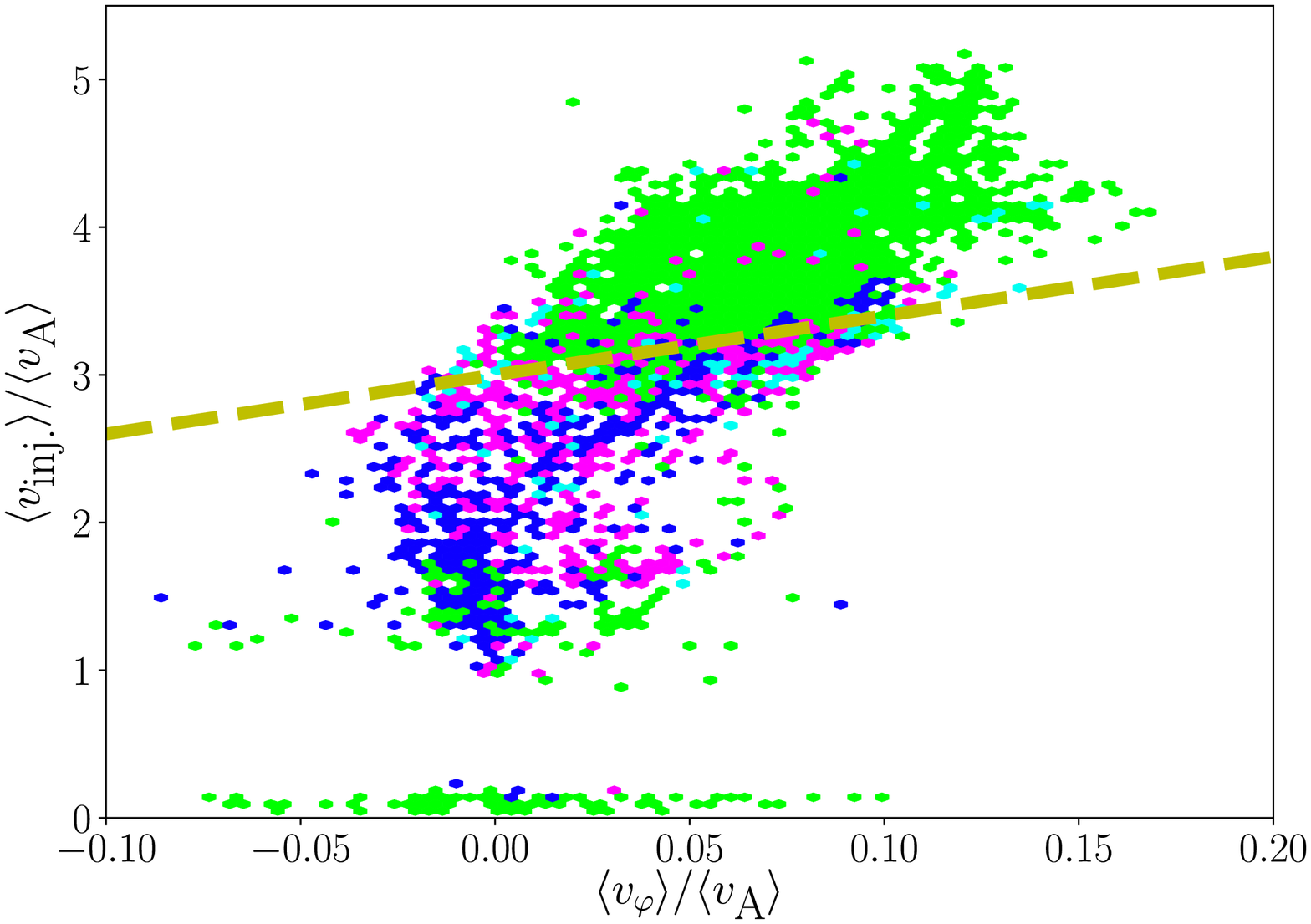}
	\caption{Normalized injection velocity versus normalized plasma toroidal velocity for TAEs. Quiescent behaviour largely confined to relatively low plasma rotation $\big(v_{\varphi} \lessapprox \frac{1}{4}(v_{\textrm{inj.}} - 3v_{\textrm{A}})\big)$.}
	\end{subfigure}
\caption{\label{fig:vinj} Plots showing differing mode character as a function of operational parameters at NSTX. \textbf{Left plots:} kink/tearing/fishbones (\SIrange{1}{30}{\kilo\hertz} modes); quiescent (green), fixed-frequency (cyan), sweeping (orange), chirping (blue), fishbone-like (magenta). \textbf{Right plots:} TAEs (\SIrange{50}{200}{\kilo\hertz} modes); quiescent (green), fixed-frequency (cyan), chirping (blue), ALEs (magenta).}
\end{figure}

\twocolumngrid

Second, we expect Alfv\'{e}nic mode structure to peak at around $q_{\textrm{min}}$,\cite{berk2001theoretical,pinches2004role,heidbrink2008basic} which we approximate to be at $\rho = \sqrt{0.5}$ (see Levinton and Yuh\cite{levinton2008motional}). Third, we extend this assumption also to the KTF modes in this analysis, however one would in reality expect the mode structure to be broader.\cite{odblom2002nonlinear} Therefore, we use the following trial function for analysis:

\begin{equation}
w(\rho) \propto \exp \left[ - \left(\dfrac{(\rho - \sqrt{0.5})^2}{2(0.1^2)} \right) \right] 
\end{equation}

such that the mode structure is approximated by a Gaussian with peak at $\rho = \sqrt{0.5}$ and standard deviation $0.1$. 

\subsection{Injection velocity}
\figref{fig:vinj} contains 4 plots showing differing mode character as a function of operational parameters at NSTX. 

\clearpage

\onecolumngrid

\begin{figure}
	\begin{subfigure}{0.49\textwidth}
	\includegraphics[width=\textwidth]{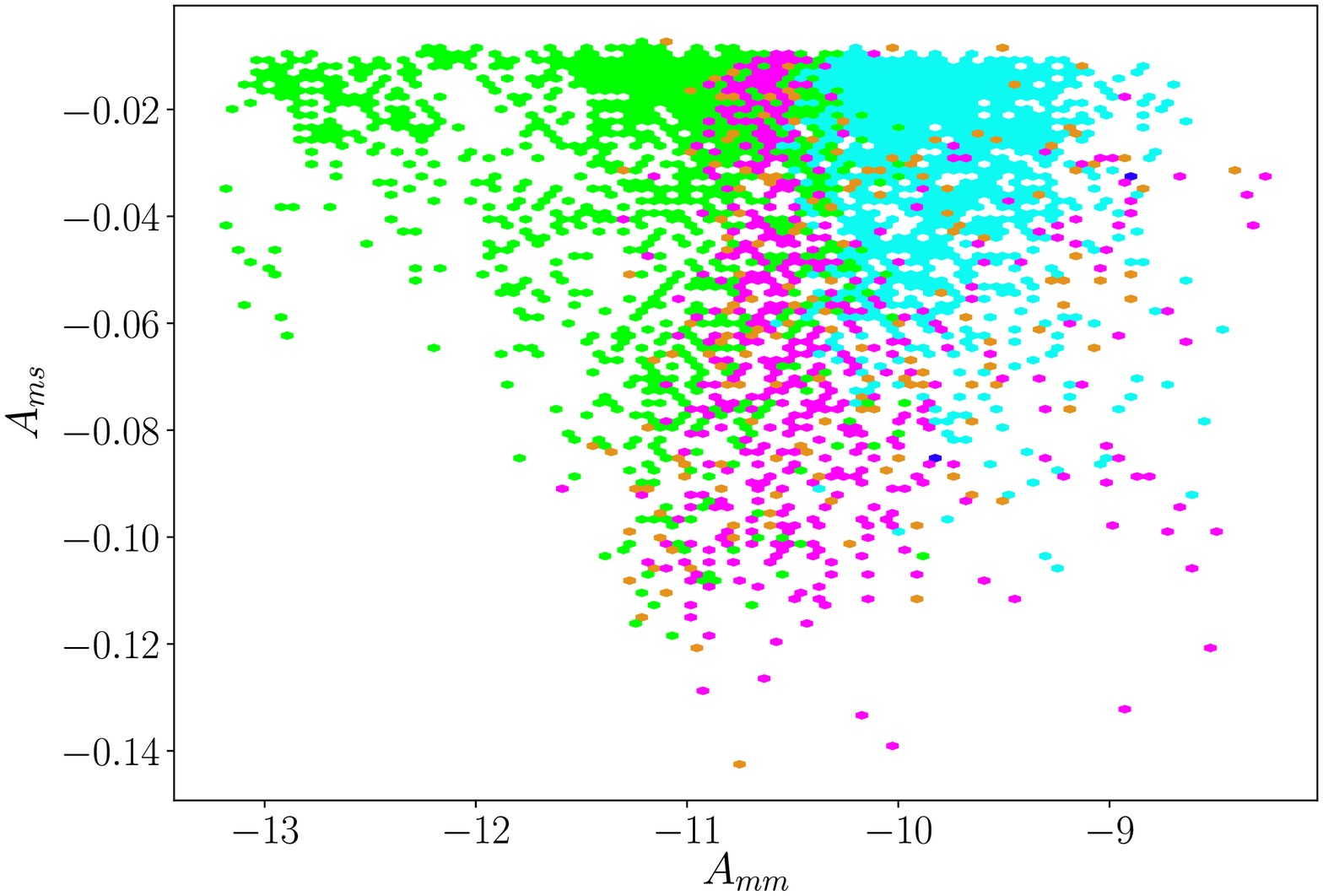}
	\end{subfigure}
	\begin{subfigure}{0.49\textwidth}
	\includegraphics[width=\textwidth]{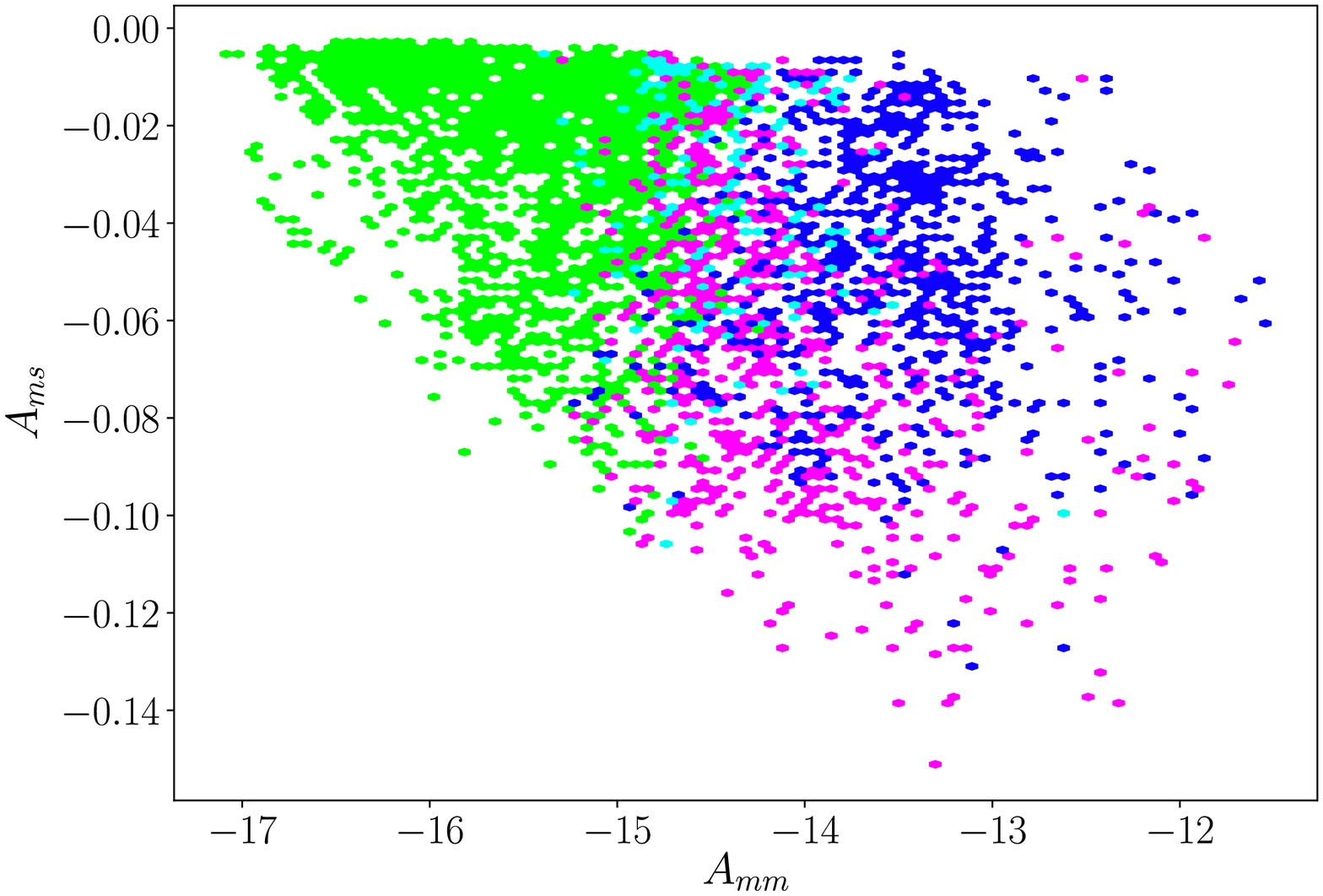}
	\end{subfigure} \\
	\begin{subfigure}{0.49\textwidth}
	\includegraphics[width=\textwidth]{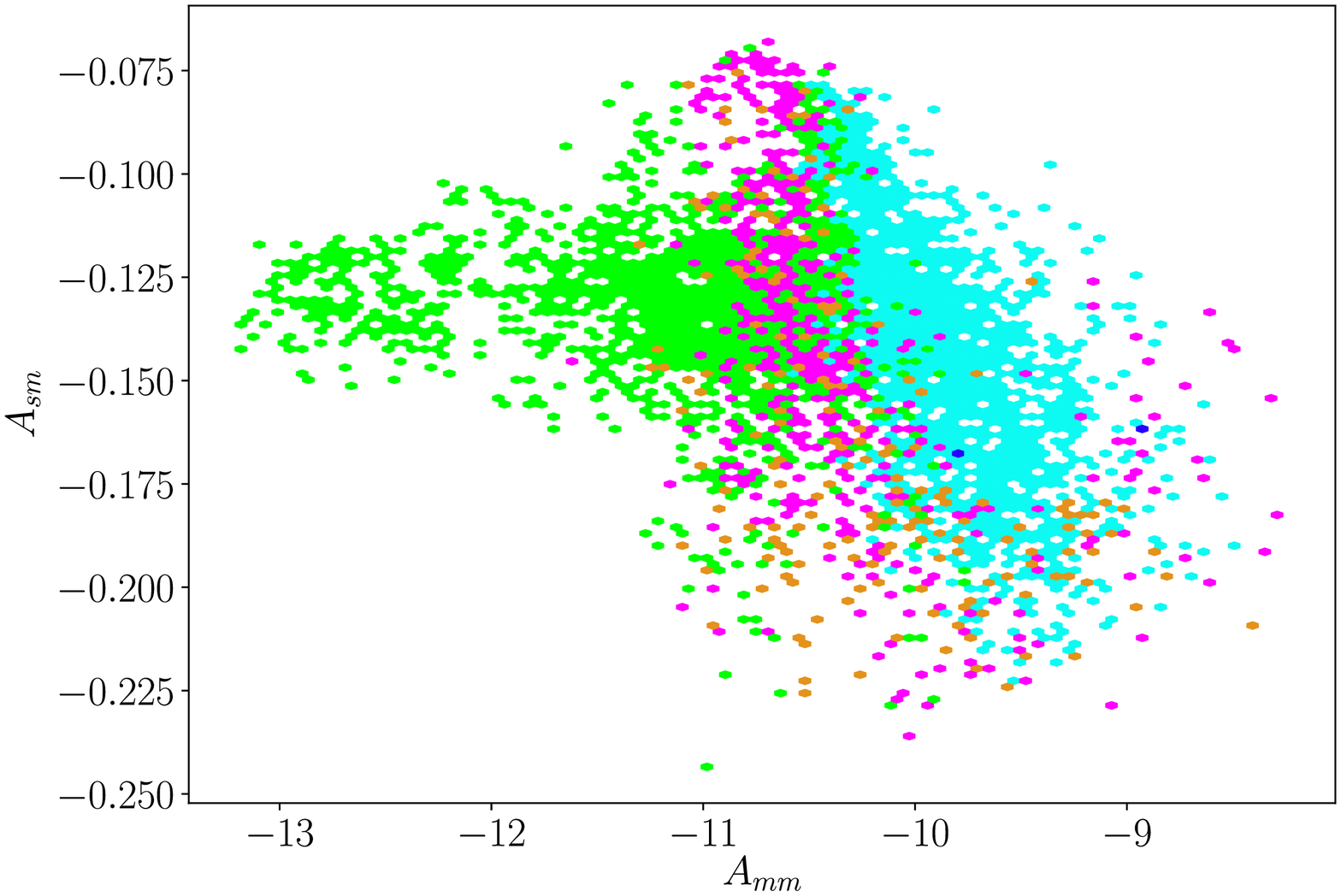}
	\end{subfigure}
	\begin{subfigure}{0.49\textwidth}
	\includegraphics[width=\textwidth]{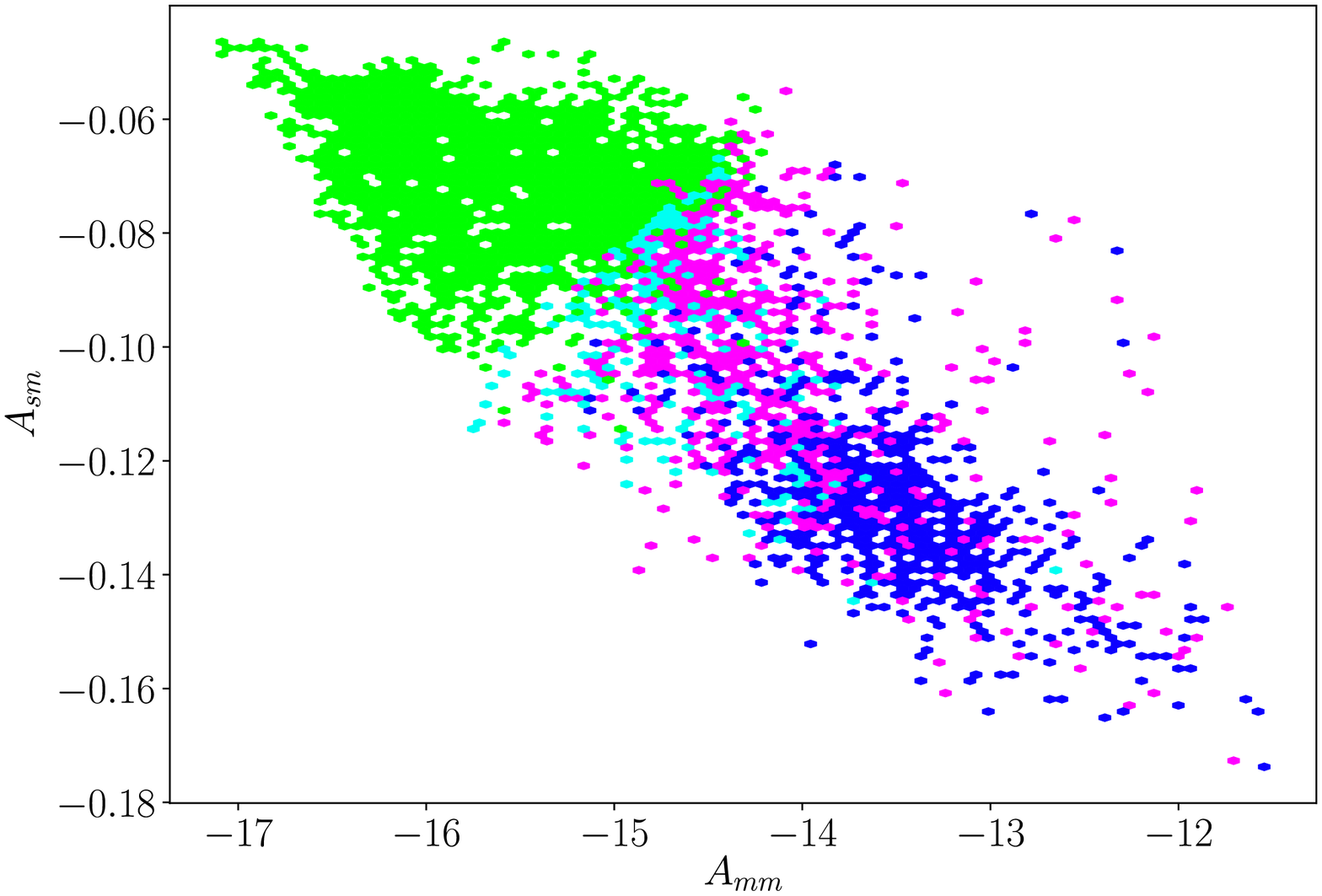}
	\end{subfigure}
\caption{\label{fig:moments} Plots showing differing mode character as a function of spectrogram moments at NSTX. \textbf{Left plots:} kink/tearing/fishbones (\SIrange{1}{30}{\kilo\hertz} modes); quiescent (green), fixed-frequency (cyan), sweeping (orange), chirping (blue), fishbone-like (magenta). Spectrogram average ($A_{\textrm{mm}}$) is a good indicator of mode character; average frequency spread ($A_{\textrm{sm}}$) and temporal intermittency ($A_{\textrm{ms}}$) are poor indicators. \textbf{Right plots:} TAEs (\SIrange{50}{200}{\kilo\hertz} modes); quiescent (green), fixed-frequency (cyan), chirping (blue), ALEs (magenta). Spectrogram average ($A_{\textrm{mm}}$) and average frequency spread ($A_{\textrm{sm}}$) are good indicators of mode character; temporal intermittency ($A_{\textrm{ms}}$) is a poor indicator.}
\end{figure}

\twocolumngrid

On the vertical axis for each plot is the normalized injection velocity $\langle v_{\textrm{inj.}} \rangle / \langle v_{\textrm{A}} \rangle$, where the Alfv\'{e}n speed is given by:

\begin{equation}
\langle v_{\textrm{A}} \rangle (t) \approx \dfrac{\langle |B| \rangle}{\sqrt{2\mu_0 \langle n_e  \rangle m_p}}
\end{equation}

where $B$ is the magnetic flux density, $n_e$ is the electron density, and $m_p$ is the proton rest mass. This approximation assumes roughly 2 atomic mass units per electron, which is suitable for the NSTX plasma in these shots (typically featuring deuterium and carbon ions). The normalized beam ion beta is defined as $\langle \beta_{\textrm{beam},i} \rangle / \langle \beta\rangle$, where $\langle \beta_{\textrm{beam},i} \rangle$ is the beam ion beta, and $\langle \beta \rangle$ is the total beta.

The left hand plots in \figref{fig:vinj} show KTF mode character (\SIrange{1}{30}{\kilo\hertz} modes); quiescent (green), fixed-frequency (cyan), sweeping (orange), chirping (blue), fishbone-like (magenta). The right hand plots in \figref{fig:vinj} show TAE mode character (\SIrange{50}{200}{\kilo\hertz} modes); quiescent (green), fixed-frequency (cyan), chirping (blue), ALEs (magenta). 

For modes in the KTF band, fixed-frequency modes are largely confined to regions of parameter space where the injection velocity is super-Alfv\'{e}nic, and the beam ion beta is relatively low. We find fixed-frequency mode behaviour for $v_{\textrm{inj.}} \gtrapprox 2 v_{\textrm{A}}$, $\beta_{\textrm{beam},i} \lessapprox 0.25 \beta$, $v_{\varphi} \gtrapprox 0.05 v_{\textrm{A}}$. Quiescence is largely observed for sub-Alfv\'{e}nic toroidal velocity, approximately given by $v_{\varphi} \lessapprox 0.05 v_{\textrm{A}}$.

For modes in the TAE band, quiescent behaviour is largely confined to regions of parameter space where the beam ion beta is relatively low. We find quiescence for $\beta_{\textrm{beam},i} \lessapprox 0.2 \beta$, $v_{\varphi} \lessapprox \frac{1}{4}(v_{\textrm{inj.}} - 3v_{\textrm{A}})$.

\clearpage

\onecolumngrid

\begin{figure}
	\begin{subfigure}{0.49\textwidth}
	\includegraphics[width=\textwidth]{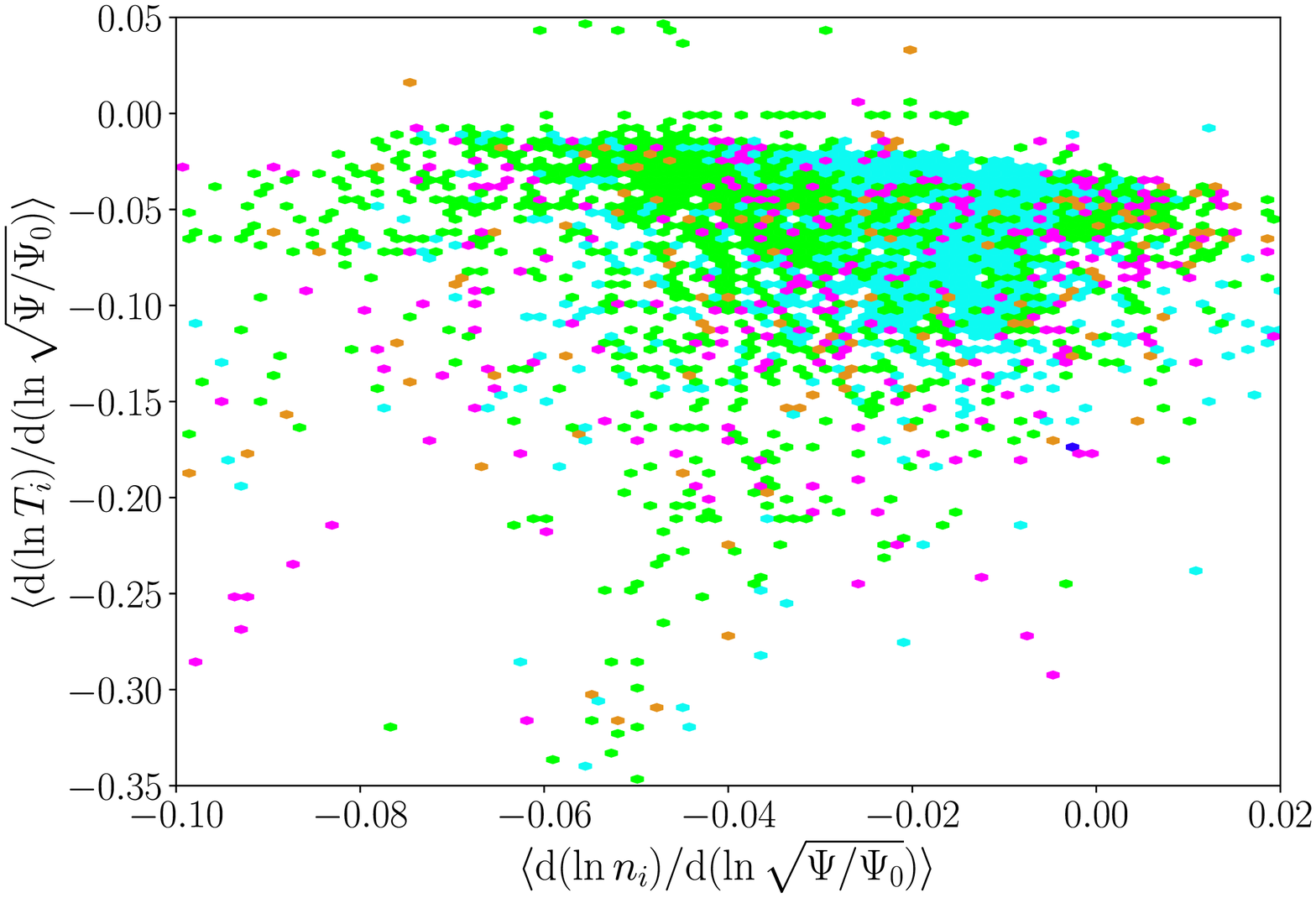}
	\end{subfigure}
	\begin{subfigure}{0.49\textwidth}
	\includegraphics[width=\textwidth]{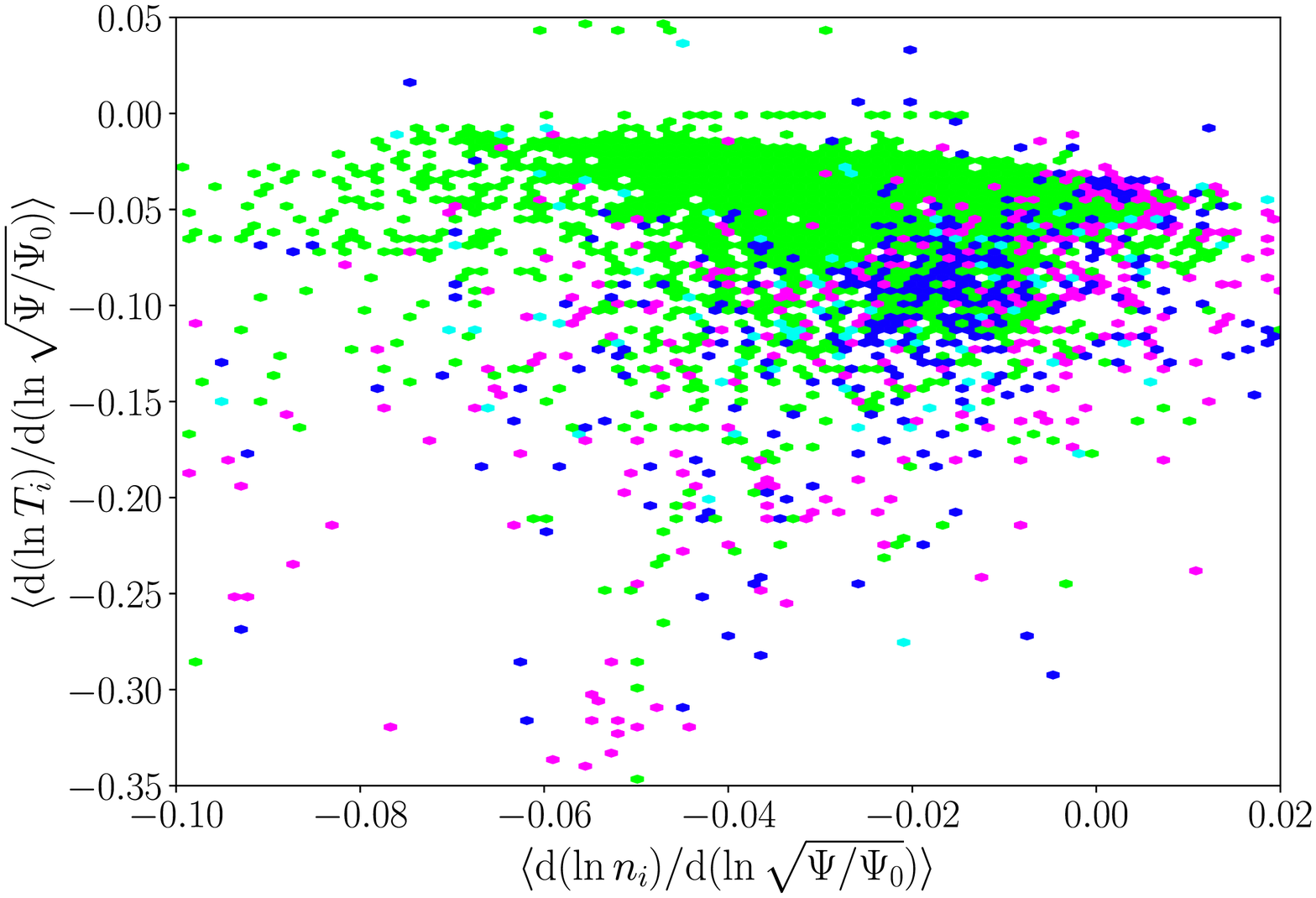}
	\end{subfigure}
\caption{\label{fig:eta} Plots showing differing mode character as a function of normalized ion temperature gradient (d$(\ln T_i)$/d$(\ln \sqrt{\Psi/\Psi_0})$) and normalized ion density gradient (d$(\ln n_i)$/d$(\ln \sqrt{\Psi/\Psi_0})$) at NSTX. \textbf{Left plot:} kink/tearing/fishbones (\SIrange{1}{30}{\kilo\hertz} modes); quiescent (green), fixed-frequency (cyan), sweeping (orange), chirping (blue), fishbone-like (magenta). Largely null result; no strong correlations observed. \textbf{Right plot:} TAEs (\SIrange{50}{200}{\kilo\hertz} modes); quiescent (green), fixed-frequency (cyan), chirping (blue), ALEs (magenta). Chirping and ALEs occur at low $\eta = (\nabla T_i / \nabla n_i)$.}
\end{figure}

\twocolumngrid

Plasma rotation can play a strong role in ideal MHD stabilisation; sub-Alfv\'{e}nic plasma rotation serves to stabilise the plasma, leading to increased quiescence in the KTF frequency band. However, the TAEs are subject to kinetic instabilities for super Alfv\'{e}nic ($v_{\textrm{inj.}} > v_{\textrm{A}}$) near the Alfv\'{e}n speed. Our results allow us to further posit for Alfv\'{e}n waves that if the injection velocity is less than the Alfv\'{e}n speed, one expects that a reversed toroidal plasma velocity would lead to decreased TAE activity.

The beam ion beta also plays a strong role in TAE destabilisation. Increased NBI power increases the kinetic drive for resonant modes; this leads to an increased likelihood for nonlinear wave-particle interaction and chirping \cite{heidbrink1995beam}. As the gradient of the fast-ion distribution determines the kinetic stability of nearby TAEs in momentum space, one expects theoretically that increased ion beam beta leads to increased TAE activity. The results from ERICSON show increased chirping and ALE activity at high beam ion beta in agreement with theory and observation.

\subsection{\label{sec:spectr}Spectrogram moments}
\figref{fig:moments} contains 4 plots showing differing mode character as a function of spectrogram moments at NSTX. The left hand plots show KTF mode character (\SIrange{1}{30}{\kilo\hertz} modes); quiescent (green), fixed-frequency (cyan), sweeping (orange), chirping (blue), fishbone-like (magenta). The right hand plots show TAE mode character (\SIrange{50}{200}{\kilo\hertz} modes); quiescent (green), fixed-frequency (cyan), chirping (blue), ALEs (magenta).

We define three simple metrics based on moments of the spectrogram signal given by \eqref{eq:signal}, taken both in frequency and time to yield a scalar. First, we define the spectrogram average $A_{mm}$ as:

\begin{equation}
A_{mm} \propto \sum\limits_{l} \sum\limits_{j} \hat{B}(f[j],t[l])
\end{equation}

$A_{mm}$ is a measure of the amplitude of magnetic fluctuations in a given frequency band. The average frequency spread $A_{sm}$ is defined as:

\begin{equation}
A_{sm} \propto \sum\limits_{l} \sqrt{\sum\limits_{j} \bigg(\hat{B}(f[j],t[l]) - \mu_f(t[l])\bigg)^2}
\end{equation}

where $\mu_f(t[l]) \propto \sum_j \hat{B}(f[j],t[l])$ is the frequency average of $\hat{B}$. $A_{sm}$ is a measure of how broadband the signal is. The intermittency $A_{ms}$ is defined as:

\begin{equation}
A_{ms} \propto \sqrt{\sum\limits_{l} \bigg(\left[\sum\limits_{j} \hat{B}(f[j],t[l])\right] - A_{mm}\bigg)^2}
\end{equation}

$A_{ms}$ is a measure of how intermittent the signal is.

For modes in the KTF band, the spectrogram average heavily dominates over the other moments; the mode character is largely invariant as a function of the average frequency spread and intermittency. For modes in the TAE band, the mode character is largely invariant as a function of the intermittency.

\subsection{Ion $\eta$}
\figref{fig:eta} contains 2 plots showing differing mode character as a function of ion temperature and pressure gradients at NSTX. The left hand plot shows KTF mode character (\SIrange{1}{30}{\kilo\hertz} modes); quiescent (green), fixed-frequency (cyan), sweeping (orange), chirping (blue), fishbone-like (magenta). The right hand plot shows TAE mode character (\SIrange{50}{200}{\kilo\hertz} modes); quiescent (green), fixed-frequency (cyan), chirping (blue), ALEs (magenta).

For modes in the KTF band, we observe a largely null result; no discernible correlation can be identified. However, for modes in the TAE band, chirping and ALEs occur at low $\eta = (\nabla T_i / \nabla n_i)$. This is consistent with observations in DIII-D \cite{duarte2017prediction}, reduced nonlinear kinetic simulations \cite{woods2018stochastic} and nonlinear gyrokinetic simulations performed for NSTX experiments \cite{duarte2018study}. Higher $\eta$ implies more drive for ion temperature gradient (ITG) modes and therefore more turbulent stochastisation of phase space. This leads to suppression of nonlinear structures, such as holes and clumps, that sustain chirping. We note, however, that each turbulent mode has its own threshold in $\eta$, hence no global threshold in $\eta$ can be identified as a well-defined transition for the chirping/fixed-frequency boundary.

\section{\label{sec:conc}Conclusions}

We have developed an ML framework to expedite the physics analysis of Alfv\'{e}n waves at NSTX. We employed Random Forest Classifiers to study correlations between plasma parameters and the frequency response of Alfv\'{e}n waves, which indicates the nature of fast ion losses.

One possible extension of this work could be towards higher frequency Alfv\'{e}n waves, such as co- and counter-propagating global Alfv\'{e}n waves (GAEs). Their frequencies have recently been shown to be highly affected by beam parameters, such as the ratio of the injection speed to the Alfv\'{e}n speed and the central pitch \cite{lestz2018energetic}, with no appreciable eigenstructure modifications. The ML framework described here can be a useful attempt to unveil possible hidden correlations in the observations.

Further work building on this analysis could aim to examine correlations in the full parameter space. Transitions such as the L-H mode transition have been partially explained in reduced parameter spaces (such as the 2D parameter space with magnetic shear $\hat{s}$ and normalised pressure gradient $\alpha$)\cite{wagner2007quarter}. This transition is shown by projecting the full parameter space onto a plane, similar to the plots we show; in reality, this transition occurs over more exotic surfaces in parameter space which are topologically challenging to analyse. Furthermore, it is entirely feasible that a coordinate transform in the parameter space would change the topology of stability boundaries - one cannot predict \emph{a priori} what is the most sensible representation of the parameter space for a given stability condition.

The work in this paper can be expanded to examine automatic dataset development (such as archival searches for shots with given mode behaviour), real-time feedback control, and real-time modeling. Recent work using ML has investigated real-time capable modeling of NSTX-U using neural networks\cite{boyer2019real}; by also employing classification data, it may be possible to further enhance the predictive capabilities of such a framework.

For both the TAE and KTF bands, very strong correlations are found between mode character and moments of the spectrograms of magnetic fluctuations in the plasma found in \secref{sec:spectr}. Accordingly, one could use these moments as features instead; one can expect that reduction to this three dimensional space yields higher accuracy at quicker computational speeds.

\section{Acknowledgments}
One of the authors (BJQW) was funded by the EPSRC Centre for Doctoral Training in Science and Technology of Fusion Energy grant EP/L01663X. VND, EDF, NN and MP acknowledge support of the US Department of Energy (DOE) under contract DE-AC02-09CH11466. This work has been carried out within the framework of the EUROfusion Consortium and has received funding from the Euratom research and training programme 2014-2018 and 2019-2020 under grant agreement No 633053. The views and opinions expressed herein do not necessarily reflect those of the European Commission. 

BJQW and VND would like to thank G. Hammett, C. Rea, and M. Van Zeeland for useful discussions. All authors would like to thank D. Boyer for insightful comments and discussions about this work.

Data used in this article is available on the ARK Dataspace: (http://arks.princeton.edu/ark:/88435/dsp01wh246w012).

\appendix
\section{Signal cross-correlation}
\label{app:cr}
If one represents the signal $V(\mathbf{r},t)$ using a Fourier transform:

\[
V(\mathbf{r},t) = \int\limits_0^{\infty} \tilde{V}(\mathbf{r},\omega) \rme^{\rmi \omega t} \dif \omega
\]

the complex Fourier spectrum $\tilde{V}(\mathbf{r},\omega)$ can be represented by using an amplitude and phase:

\[
\tilde{V}(\mathbf{r},\omega) = |\tilde{V}|(\mathbf{r},\omega) \rme^{\rmi \phi(\mathbf{r},\omega)}
\]

Then, let us define the shorthand $A(\mathbf{r} = \mathbf{r}_j) = A_j$. Then, for two signals $\{V_1(\mathbf{r}_1), V_2(\mathbf{r}_2)\}$, we define:

\[
\delta \tilde{V}(\mathbf{r}_1,\mathbf{r}_2,\omega)  = \tilde{V}_2 - \tilde{V}_1
\]

such that $\delta \tilde{V}$ is the difference in spectra. We then define the following power spectra:

\begin{align*}
\tilde{S}_{\textrm{cr.}}(\mathbf{r}_1,\mathbf{r}_2,\omega) = \tilde{V}_1^* \tilde{V}_2 &;& \tilde{S}_{\textrm{avg.}}(\mathbf{r}_1,\mathbf{r}_2,\omega) = \left(\dfrac{\tilde{V}_1 + \tilde{V}_2}{2}\right)^2
\end{align*}

where $\tilde{S}_{\textrm{cr.}}$ is the Fourier transform of the cross-correlation, and $\tilde{S}_{\textrm{avg.}}$ is the square of the Fourier transform of the average. It is trivial to show in the limit that $|\delta \tilde{V}| \ll |\tilde{V}_1|$, to first order in $\delta \tilde{V}$ the power spectra are identical if $\tilde{V}_1 \equiv \tilde{V}_1^*$. Consequently, by separating into amplitude and phase, with some abuse of notation one can define a correlation length spectrum $L_{\textrm{cr.}}(\omega)$:

\begin{align*}
\forall |\mathbf{r}_2 - \mathbf{r}_1| \leq L_{\textrm{cr.}}(\omega) : \left\{
    \begin{array}{r l} 
    |\delta |\tilde{V}|(\mathbf{r}_1,\mathbf{r}_2,\omega)| &\ll |\tilde{V}_1|, |\tilde{V}_2| \\
    |\delta \phi(\mathbf{r}_1,\mathbf{r}_2,\omega)| &\ll \pi
    \end{array}\right.
\end{align*}

This typically requires that the change in amplitude is small, but also that the two signals are suitably correlated in phase. As an example, if $V_2 = - V_1$, such that $|\delta \phi(\mathbf{r}_1,\mathbf{r}_2,\omega)| = \pi$, then $|\delta |\tilde{V}|| \ll |\tilde{V}_1|, |\tilde{V}_2|$ is satisfied, but:

\[
\tilde{S}_{\textrm{avg.}} = 0
\]

giving a significant difference in power spectra, as the phase difference is large.

\small
\bibliographystyle{unsrt}
\bibliography{biblio}

\end{document}